\documentclass[12pt]{article}

\global\arraycolsep=1pt
\usepackage{amsmath}
\usepackage{amssymb}
\newcommand{\appMac}{A}
\newcommand{\appProof}{B}
\newcommand{\appInteger}{C}
\newcommand{\appSelberg}{D}
\newcommand{\appQone}{E}
\newcommand{\appNotation}{F}
\newcommand{\be}{\begin{equation}}
\newcommand{\ee}{\end{equation}}
\newcommand{\ba}{\begin{eqnarray}}
\newcommand{\ea}{\end{eqnarray}}
\newcommand{\ds}{\displaystyle}
\renewcommand{\theequation}{\thesection.\arabic{equation}}
\newcommand{\Section}{\setcounter{equation}{0} \section}
\newcommand{\Appendix}[2]{
\renewcommand{\theequation}{#1.\arabic{equation}}\setcounter 
{equation}{0}
\renewcommand{\thesubsection}{#1.\arabic{subsection}}\setcounter 
{subsection}{0}
\section*{Appendix #1: #2}
}
\renewcommand{\thefootnote}{\fnsymbol{footnote}}
\def\Xint#1{\mathchoice
   {\XXint\displaystyle\textstyle{#1}}%
   {\XXint\textstyle\scriptstyle{#1}}%
   {\XXint\scriptstyle\scriptscriptstyle{#1}}%
   {\XXint\scriptscriptstyle\scriptscriptstyle{#1}}%
   \!\int}
\def\XXint#1#2#3{{\setbox0=\hbox{$#1{#2#3}{\int}$ }
   \vcenter{\hbox{$#2#3$ }}\kern-.5\wd0}}

\def\dashint{\Xint-}
\def\ddelta{\delta}
\newcommand{\bC}{{\mathbb C}}
\newcommand{\bN}{{\mathbb N}}
\newcommand{\bZ}{{\mathbb Z}}
\newcommand{\cF}{{\cal F}}
\newcommand{\cN}{{\mathcal N}}
\newcommand{\cS}{{\mathcal S}}
\newcommand{\cO}{{\cal O}}
\newcommand{\cW}{{\cal W}}
\newcommand{\proof}{\noindent{\it Proof.\hskip10pt}} 
\newcommand{\qed}{\hfill\fbox{}}
\newcommand{\ha}{{1\over2}}
\newcommand{\Exp}[1]{\exp\left\{#1\right\}}
\newcommand{\deldel}[1]{ {\frac{\partial~}{\partial #1}} }

\newcommand{\BULLET}{\bullet}
\newcommand{\NP}{{\rlap{\raise.4ex\hbox{$\scriptscriptstyle\BULLET$}}
               \lower.4ex\hbox{$\scriptscriptstyle\BULLET$}}}
\newcommand{\twoPiI}{2\pi i}
\newcommand{\overtwoPiI}{\over 2\pi i}

\newcommand{\CF}[1]{ \left\langle\!\left\langle\, #1 \,\right\rangle\!\right\rangle }

\newcommand{\qpint}[2]{ \left[\, #1 \,\right]_{#2} }

\newcommand{\CT}[2]{{\rm CT}_{\{#1\}} #2 }

\newcommand{\qHyperGeo}[7]{
{}_{#1} \varphi_{#2}
\left[
{ #5 \atop #6 } 
;#3,#7
\right]
}

\newcommand{\qtHyperGeo}[7]{
{}_{#1} \varphi_{#2}^{(#3,#4)}
\left[
{ #5 \atop #6 } 
;#7
\right]
}

\newcommand{\betaHyperGeo}[6]{
{}_{#1} \varphi_{#2}^{#3}
\left[
{ #4 \atop #5 } 
;#6
\right]
}

\newcommand{\qW}{{$q$-${\cal W}_\nn $}}
\newcommand{\W}{{${\cal W}_\nn $}}

\newcommand{\tcW}{{\cal\widetilde W}}
\newcommand{\cLa}{{\hat \Lambda}}
\newcommand{\tcLa}{{\widetilde \Lambda}}

\newcommand{\xp}[1]{x\!\left[\, #1\, \right]}

\newcommand{\dualvee}{'}

\newcommand{\qt}{\xi}
\newcommand{\tN}{u}

\newcommand{\gs}{{g_s}}

\newcommand{\tz}{{\tilde z}}
\newcommand{\tp}{{\tilde p}}
\newcommand{\tilr}{{\tilde r}}
\newcommand{\tilk}[1]{{\tilde #1}}

\newcommand{\alphalk}[2]{\alpha}

\newcommand{\nn}{N}
\newcommand{\mm}{r}
\newcommand{\massLambda}[2]{\Lambda^{#1}_{#2}}
\newcommand{\massQ}[1]{Q^{#1}}
\newcommand{\pQ}{Q^+}
\newcommand{\mQ}{Q^-}

\newcommand\qWDelta{\Delta^{qW}}
\newcommand\qHyperDelta{\Delta^{qH}}
\newcommand\MacDelta{\Delta^{Mac}}	
\newcommand\rb{\sqrt{\beta}}
\newcommand\rbi{-{1\over\sqrt{\beta}}}

\newcommand\VO{V}

\newcommand\IP[3] {\langle #1,#2 \rangle}
\newcommand\MacIP[3]{\langle #1,#2 \rangle_{#3}'} 
\newcommand\qWIP[3]{\langle #1,#2 \rangle_{#3}''} 

\newcommand\PIP[1] {\langle\, #1\, \rangle}
\newcommand\MacPIP[2]{\langle\, #1\, \rangle_{#2}'} 
\newcommand\qWPIP[2]{\langle\, #1\, \rangle_{#2}''} 

\newcommand\PIPbk[1] {\langle\, #1\, \rangle}
\newcommand\MacPIPbk[2]{\langle\, #1\, \rangle_{#2}'} 
\newcommand\qWPIPbk[2]{\langle\, #1\, \rangle_{#2}''}

\newcommand\qWC[2]{\widetilde C_{#1}^{#2}} 

\newcommand{\h}[2]{h^{#1}_{#2}}		
\newcommand{\Qh}[1]{Q_h^{#1}} 		
\newcommand{\al}[2]{\alpha^{#1}_{#2}}	
\newcommand{\Qal}[1]{Q_\alpha^{#1}} 	
\newcommand{\Lam}[2]{\Lambda^{#1}_{#2}}	
\newcommand{\QLa}[1]{Q_\Lambda^{#1}} 	

\newcommand{\Thalpha}[3]{A^{#1,#2}(#3)}
\newcommand{\Talphah}[3]{(A^{-1})^{#1,#2}(#3)}
\newcommand{\ThLambda}[3]{B^{#1,#2}(#3)}
\newcommand{\TLambdah}[3]{(B^{-1})^{#1,#2}(#3)}
\newcommand{\TalphaLambda}[3]{C^{#1,#2}(#3)}
\newcommand{\TLambdaalpha}[3]{(C^{-1})^{#1,#2}(#3)}

\newcommand{\sfrac}[2]{{\textstyle \frac{#1}{#2}}}

\newcommand{\bosonIP}[3]{ ({#1}\cdot{#2}{#3}) }

\newcommand{\ch}[1]{{\bosonIP{\vec h^{#1}}{\cphi}{ } }}
\newcommand{\tch}[1]{{\bosonIP{\vec h^{#1}}{\tcphi}{ } }}

\newcommand{\Wcl}{W_{c\ell}}
\newcommand{\cWcl}{{\cal W}_{c\ell}}
\newcommand{\tcWcl}{{\widetilde{\cal W}}_{c\ell}}

\newcommand{\cphi}{\partial_z\vec{\hat\phi}(z)}
\newcommand{\tcphi}{\partial_z\vec{\widetilde\phi}(z)}

\def\boxline#1{\vbox{\hrule\hbox{\vrule\vbox{#1}\vrule}\hrule}}
\def\boxNW#1{\vbox{\hrule\hbox{\vrule\vbox{#1}}}}
\def\boxES#1{\vbox{\hbox{\vbox{#1}\vrule}\hrule}}
\def\Between#1#2#3#4{ 
\raise-#1mm\vbox to#1mm{\hsize #2mm \vbox{\vskip #3mm\centerline{#4} } }
}
\def\Square#1#2#3#4{ 
\raise-#1mm\boxline{
\vbox to#1mm{\hsize #1mm \vbox{\vskip #2mm\noindent\hskip4pt {#3} 
			       \vskip-#2mm\vskip-8pt\centerline{#4} }} }
}
\def\Young#1#2#3#4#5#6#7#8#9{ 
\raise-#9mm\boxNW{\vbox to#1mm{\hsize#6mm 
	\vbox{\vskip#8mm\noindent\hskip#8mm$\;\lambda$} }}
\kern-#6mm 
\raise-#1mm\boxES{\vbox to #5mm{\hsize #7mm $ $}}\kern-.4pt
\raise-#2mm\boxES{\vbox to #5mm{\hsize #7mm $ $}}\kern-.4pt
\raise-#3mm\boxES{\vbox to #5mm{\hsize #7mm $ $}}\kern-.4pt
\raise-#4mm\boxES{\vbox to #5mm{\hsize #7mm $ $}}\kern-.4pt
\raise-#5mm\boxES{\vbox to #5mm{\hsize #7mm $ $}}
}
\def\Galilei{
  \Young{10}8642{15}32{9.9}
\Between{10}{20}3{$\longmapsto$}
 \Square{14}7{$r$}{$s$} \kern-.4pt  
  \Young{10}8642{15}32{9.9}
\Between{10}{10}{10}{\quad .}
}
\def\generalYoung{
\vskip.25cm
\noindent
\makebox[  3cm]{ }
\makebox[  2cm]{$s_1$}\hskip-.4pt
\makebox[1.7cm]{$s_2$}
\makebox[1.4cm]{ }
\makebox[1.4cm]{$s_{N-2}$}\hskip-.35pt
\makebox[1.3cm]{$s_{N-1}$}
\hfill\break
 \makebox[  3cm][r]{$\hfill\lambda=$}
\framebox[  2cm][l]{\rule[  -1cm]{0cm}{  2cm}$r_1$}\hskip-.4pt
\framebox[1.7cm][l]{\rule[-0.7cm]{0cm}{1.7cm}$r_2$}
 \makebox[1.4cm]			    {\raisebox{.25cm}{$\cdots\cdots$}}
\framebox[1.4cm][l]{\rule[-0.4cm]{0cm}{1.4cm}\raisebox{.25cm}{$r_{N-2}$}
							     }\hskip-.4pt
\framebox[1.3cm][l]{\rule[-0.2cm]{0cm}{1.2cm}\raisebox{.25cm}{$r_{N-1}$}}
\makebox[1cm][r]{.}
\vskip.3cm
}

\begin{document}


\begin{titlepage}
\begin{flushright}
{April 28, 2010 }
\end{flushright}
\vspace{0.5cm}
\begin{center}
{\Large \bf
Five-dimensional AGT Relation \\ 
\vskip-3pt
and \\ 
\vskip6pt
the Deformed $\beta$-ensemble
}
\vskip10mm
{
{
Hidetoshi Awata}${}^\star$ 
and 
{
Yasuhiko Yamada}${}^\dagger$
}
\vskip 10mm
{\baselineskip=14pt
\it
${}^\star$%
Graduate School of Mathematics \\
Nagoya University, Nagoya,  464-8602, Japan\\
\vskip12pt
${}^\dagger$%
Department of Mathematics, Faculty of Science \\
Kobe University, Hyogo 657-8501, Japan
}
\end{center}
\vskip8mm

\begin{abstract}
We discuss an analog of the AGT relation in five dimensions.
We define a $q$-deformation of the $\beta$-ensemble 
which satisfies {\qW} constraint.
We also show a relation with 
the Nekrasov partition function of 
5D $SU(\nn)$ gauge theory with $N_f=2\nn$. 
\end{abstract}
\end{titlepage}


\renewcommand{\thefootnote}{\arabic{footnote}} \setcounter 
{footnote}{0}


\Section{Introduction}


In \cite{rf:Alday-Gaiotto-Tachikawa}, 
Alday Gaiotto and Tachikawa discovered
remarkable relations between 
the 4D $\cN=2$ super conformal gauge theories 
and the 2D Liouville CFT.
Some explanations have been addressed from
$\beta$-ensemble (generalized matrix model)  
\cite{rf:Harris}\cite{rf:AMOS:Note}
in \cite{rf:Dijkgraaf-Vafa}--\cite{rf:MM}.

In the pure $SU(2)$ case, the AGT relation \cite{rf:Gaiotto} 
between the instanton part of the partition functions of the gauge theory 
and correlation functions of the Virasoro algebra 
is extended naturally to 5D in \cite{rf:Awata-Yamada}.%
\footnote{
The recursion relation for the 5D partition functions 
is derived recently in \cite{rf:Yanagida}.
}{ }
The instanton counting
\cite{rf:Nekrasov}--
\cite{rf:Nakajima-Yoshioka} %
of the 5D gauge theory 
\cite{rf:Nakajima-Yoshioka:05} 
can be viewed as a $q$-analog of 4D cases
\cite{rf:Awata-Kanno}--
\cite{rf:Awata-Kanno:08},
and there also exists a natural $q$-deformation of the Virasoro/$\cW_\nn$ algebra
\cite{rf:SKAO}--%
\cite{rf:AKOS}.

In this article, 
we will study a 5D extension of the AGT relation with $N_f=2\nn$.
The $A_{\nn-1}$ type quiver matrix model (the ITEP model) \cite{rf:KMMM}
was generalized as a $\beta$-ensemble \cite{rf:Harris}
satisfying the ${\cal W}_\nn$ constraint by \cite{rf:AMOS:Note}.
Under the strategy of \cite{rf:AMOS:Note},
we will introduce $q$-deformed $\beta$-ensemble 
which automatically satisfies {\qW} constraint
and show a relation with 
the 5D Nekrasov partition function of 
$SU(\nn)$ gauge theory with $N_f=2\nn$. 


This paper is organized as follows:
In section 2,
we start with recapitulating the result of the {\qW} algebra
and also define primary fields.
In section 3, 
we introduce $q$-deformed $\beta$-ensemble 
which automatically satisfies {\qW} constraint.
Section 4 
deals with the $\nn=2$ case.
Finally in section 5,
we explain a reduction of the 5D Nekrasov partition function 
to the $q$-hypergeometric function 
and show a coincidence with 
the partition function of our $q$-deformed $\beta$-ensemble. 
Appendix A 
contains a definition of the Macdonald polynomial and several useful formulas.
Proof of the key equation is shown 
in appendix B.
Relations with 
the Kaneko's integral formula and 
the Jackson integral formulas are given 
in appendices C and D, respectively.
In appendix E, 
we review the 4D case.
Appendix F 
is devoted to a list of notations for bosons.


\noindent{\bf Notation.}
%
%
Let 
${\qpint np} := 
( p^{{n\over 2}}-p^{-{n\over 2}} )/( p^{\ha}-p^{-\ha} )$.
Parameters are 
$q:=e^{\hbar /\sqrt\beta}=e^{\gs R}$, 
$t:= q^\beta
=e^{\hbar \sqrt\beta} =e^{\gs \beta R}
$, 
$p:= q/t=e^{-\hbar( \sqrt\beta-1/\sqrt\beta)}$, 
$\tN := t^\gamma$
and $v:=(q/t)^\ha$.
We will use the same letter $p$ also for 
the set of power sums $p:=(p_1,p_2,\cdots)$,
but this 
appears only at 
$P_\lambda({\xp p})$ or $Z_2(p)$.

\Section{Quantum deformation of $\cW_\nn $ algebra} 


We start with recapitulating the results of the {\qW} algebra
\cite{rf:Feigin-Frenkel} \cite{rf:AKOS}
and define primary fields.
%
%
%

\subsection{Bosons} 


We use three kinds of basis for bosons.
First we define fundamental bosons ${\h in}$ and ${\Qh i}$ 
for $i=1,2,\cdots,\nn $ and  $n\in\bZ$ such that
\ba
[{\h in},{\h jm}] &=& 
{1\over n}(q^{{n\over 2}}-q^{-{n\over 2}})(t^{{n\over 2}}-t^{-{n\over 2}})
{ {\qpint{\delta_{ij}\nn -1}{p^n}} \over {\qpint{\nn }{p^n}} } 
p^{{n\over 2} \nn {\rm sgn}(j-i)}
	\ddelta_{n+m,0},\cr
[{\h in},{\Qh j}] &=& 
\left(\delta_{ij} - {1\over \nn }\right)\ddelta_{n,0},
\qquad
[{\Qh i},{\Qh j}] =0,
\qquad
\sum_{i=1}^\nn p^{in} {\h in} = 0,
\qquad
\sum_{i=1}^\nn {\Qh i} =0
~~
\ea
with $q$, $t:= q^\beta\in\bC$, $p:= q/t$,
${\qpint np} := 
( p^{{n\over 2}}-p^{-{n\over 2}} )/( p^{\ha}-p^{-\ha} )$
and 
${\rm sgn}(i):=1$, $0$ or $-1$ for $i>0$, $i=0$ or $i<0$, respectively.
Here $[A,B]:=AB-BA$.
This bosons correspond to the weights $\vec h_i$ of the vector representation
whose inner product is $(\vec h_i\cdot \vec h_j)= \delta_{ij} -1/\nn $.
This algebra is invariant under the following involutions: 
$\omega_\pm^2 = 1$,
\ba
\omega_+
&:&
\quad
\sqrt\beta\mapsto 1/\sqrt\beta,
\quad
(q,t) \mapsto (t,q),
\quad\hskip22pt
{\h in} \mapsto {\h {\nn-i+1}n},
\quad
{\Qh i}\mapsto {\Qh {\nn-i+1}},
\label{eq:Involqttq}%
\\
\omega_-
&:&
\quad
\sqrt\beta\mapsto -\sqrt\beta,
\quad\hskip2pt
(q,t) \mapsto (q^{-1},t^{-1}),
\quad
{\h in}\mapsto{\h {\nn-i+1}n},
\quad
{\Qh i}\mapsto {\Qh {\nn-i+1}}.
\label{eq:Involqqinv}%
\ea

%
Next let us introduce root type bosons 
${\al an} := {\h an}-{\h {a+1}n}$ and 
${\Qal a} := {\Qh a}-{\Qh {a+1}}$ for $a=1,2,\cdots,\nn -1$.
Then they satisfy
\ba
[{\al an},{\al bm}] &=& 
{1\over n}(q^{{n\over 2}}-q^{-{n\over 2}})(t^{{n\over 2}}-t^{-{n\over 2}})
{\TalphaLambda ab{p^n}}
	\ddelta_{n+m,0},\cr
[{\al an},{\Qal b}] &=& 
{\TalphaLambda ab1}
\ddelta_{n,0},
\qquad
[{\Qal a},{\Qal b}] =0
\ea
and 
\ba
[{\h in},{\al bm}] &=& 
{1\over n}(q^{{n\over 2}}-q^{-{n\over 2}})(t^{{n\over 2}}-t^{-{n\over 2}})
{\ThLambda ib{p^n}}
	\ddelta_{n+m,0},
\cr
[{\h in},{\Qal b}] &=& 
{\ThLambda ib1}
\ddelta_{n,0}
=
[{\al bn},{\Qh i}],
\qquad
[{\Qh i},{\Qal b}] = 0.
\ea
Here
\ba
{\ThLambda ibp} 
&:=&
p^{\ha}\delta_{i,b} - p^{-\ha}\delta_{i-1,b},
\cr
{\TalphaLambda abp}
&:=&
- p^{-\ha}\ddelta_{a-1,b}
+{\qpint 2p}\ddelta_{a,b} 
- p^{\ha}\ddelta_{a+1,b}.
\ea
Note that 
$[{\h an} + p^n {\h {a+1}n}, {\al am}] = 0$.

%
Finally we define 
weight type bosons 
${\Lam an} := \sum_{b=1}^a {\h bn} p^{(b-a-\ha)n}$ and 
${\QLa a} := \sum_{b=1}^a {\Qh b}$ for $a=1,2,\cdots,\nn -1$.
Note that
${\h an} = p^{{n\over 2}}{\Lam an}-p^{-{n\over 2}}{\Lam {a-1}n}$
with 
${\Lam 0n} := 0$. 
Then they satisfy
\ba
[{\al an},{\Lam bm}] &=& 
{1\over n}(q^{{n\over 2}}-q^{-{n\over 2}})(t^{{n\over 2}}-t^{-{n\over 2}})
\ddelta_{a,b}\ddelta_{n+m,0},
\cr
[{\al an},{\QLa b}]
&=& 
\ddelta_{a,b}\ddelta_{n,0}
=
[{\Lam bn},{\Qal a}],
\qquad
[{\Qal a},{\QLa b}] =0,
\ea
\ba
[{\h in},{\Lam bm}] &=& 
{1\over n}(q^{{n\over 2}}-q^{-{n\over 2}})(t^{{n\over 2}}-t^{-{n\over 2}})
{\Thalpha ib{p^{-n}}}
\ddelta_{n+m,0},
\cr
[{\h in},{\QLa b}]
&=& 
{\Thalpha ib1}
\ddelta_{n,0}
=
[{\Lam bn},{\Qh i}] ,
\qquad
[{\Qh i},{\QLa b}] =0
\ea
and
\ba
[{\Lam an},{\Lam bm}] &=& 
{1\over n}(q^{{n\over 2}}-q^{-{n\over 2}})(t^{{n\over 2}}-t^{-{n\over 2}})
{\TLambdaalpha ab{p^n}}
\ddelta_{n+m,0},
\cr
[{\Lam an},{\QLa b}] &=& 
{\TLambdaalpha ab1}
\ddelta_{n,0},
\qquad
[{\QLa a},{\QLa b}] =0.
\ea
Here
\ba
{\Thalpha ibp}
&:=&
{ { \qpint{ \nn \theta(i\leq b) - i }p } \over { \qpint \nn p } }
p^{\ha( b-\nn \theta(i>b) )},
\cr
{\TLambdaalpha abp}
&=&
{ { \qpint{ \min (a,b) }p }{ \qpint{ \nn -\max (a,b) }p } 
\over { \qpint \nn p } }
p^{{b-a\over 2}}
\ea
with $\theta(P):= 1$ or $0$ 
if the proposition $P$ is true or false, respectively.
%
%
Note that
by $\omega_\pm$,
\ba
\omega_\pm
&:&
\quad
{\al an} \mapsto -{\al {\nn-a}n},
\quad\hskip50pt
{\Qal a}\mapsto -{\Qal {\nn-a}},
\cr
\omega_\pm
&:&
\quad
{\Lam an} \mapsto -{\Lam {\nn-a}n} p^{(a-\nn-\ha)n},
\quad
{\QLa a}\mapsto -{\QLa {\nn-a}}.
\ea


\subsection{{\qW} algebra}



Let us define fundamental vertices $\Lambda_i(z)$ and 
{\qW} generators $W^i(z)$ for $i=1,2,\cdots,\nn $ as follows:
\ba
\Lambda_i(z) &:= & \NP \Exp{ \sum_{n\neq 0}{\h in} z^{-n} }\NP \,
q^{\rb{\h i0}} p^{{\nn +1\over 2}-i},\cr
W^i(zp^{1-i\over 2}) &:= & \sum_{1\leq j_1<\cdots<j_i\leq \nn }
\NP \,\Lambda_{j_1}(z) \Lambda_{j_2}(zp^{-1}) \cdots \Lambda_{j_i}(zp^{1-i})\,\NP
\ea
and $W^0(z):= 1$. 
Here $\NP *\NP $ stands for the usual bosonic normal ordering such that
the bosons ${\h in}$ with non-negative mode $n\geq 0$ are in the right. 
Note that
\be
W^\nn (zp^{1-\nn \over 2}) =
\NP \,\Lambda_1(z) \Lambda_2(zp^{-1}) \cdots \Lambda_\nn (zp^{1-\nn })\,\NP  = 1.
\ee
These generators are obtained from the following quantum Miura transformation:
\be
\sum_{i=0}^\nn (-1)^i W^i(zp^{1-i\over 2}) p^{(\nn -i)D_z}
=
\NP 
\left(p^{D_z} - \Lambda_1(z)\right) 
\left(p^{D_z} - \Lambda_2(zp^{-1})\right) 
\cdots
\left(p^{D_z} - \Lambda_\nn (zp^{1-\nn })\right)
\NP
\label{eq:qMiura}%
\ee
with $D_z := z{\partial\over \partial z}$.
Remark that $p^{D_z}$ is the $p$-shift operator such that
$p^{D_z} f(z) = f(pz)$.
The mode $n$ generator $W^i_n$ 
is defined by $W^i(z) =: \sum_{n\in\bZ} W^i_n z^{-n}$.

\newpage
\subsection{Screening currents}


%
%
By using root type bosons we define screening currents $S^a_\pm(z)$ as follows:
\be
S^a_\pm(z) := 
\NP \Exp{ \mp\sum_{n\neq0}{{\al an}\over \qt_\pm^{n\over 2} - \qt_\pm^{-{n\over 2}} } 
z^{-n} }\NP \,
e^{\pm\rb^{\pm1}{\Qal a}} z^{\pm\rb^{\pm1}{\al a0}},
\qquad
\qt_+ = q,
\quad
\qt_- = t. 
\label{eq:SCDef}%
\ee
Note that the Langlands duality
$\omega_- \omega_+ S^a_+(z) = S^a_-(z)$.
We denote the negative mode part of $S^a_\pm(z)$ by
$(S^a_\pm(z))_- := 
\Exp{ \mp\sum_{n<0}{{\al an}\over\qt_\pm^{n\over 2} - \qt_\pm^{-{n\over 2}} } z^{-n} }
$.
The screening currents satisfy
\ba
&&\hspace{-7mm}
\left[\,\NP \left(p^{D_z} 
- \Lambda_1(z)\right)\left(p^{D_z}
- \Lambda_2(zp^{-1})\right)\cdots
\left(p^{D_z} - \Lambda_\nn (zp^{1-\nn })\right)\NP  \,, 
S^a_\pm(w)\right]\cr
&&
= (q^\ha-q^{-\ha})(t^\ha-t^{-\ha}){d\over d_{\qt_\pm}w}  
\,\NP \left(p^{D_z} 
- \Lambda_1(z)\right)\cdots\left(p^{D_z} 
- \Lambda_{a-1}(zp^{2-a})\right) 
\cr
&&\hspace{5mm}\times
w \ddelta\left({w\over z}p^{a-1}\right) A^a_\pm(w) p^{D_z}
\left(p^{D_z} - \Lambda_{a+2}(zp^{-1-a})\right)\cdots
\left(p^{D_z} - \Lambda_\nn (zp^{1-\nn })\right) \NP
\ea
with
\be
A^a_\pm(w) :=
\NP \Exp{ \sum_{n\neq0}
{ \qt_\pm^{\pm{n\over 2}} {\h {a+1}n} - \qt_\pm^{\mp{n\over 2}}{\h an} \over 
  \qt_\pm^{\pm{n\over 2}} - \qt_\pm^{\mp{n\over 2}}}
w^{-n} }\NP \,
e^{\pm\rb^{\pm 1}{\Qal a}} w^{\pm\rb^{\pm 1}{\al a0}} 
\qt_\pm^{\ha\rb^{\pm 1}( {\h{a}0}+{\h{a+1}0} ) } 
p^{{\nn \over 2}-a}
\ee
and 
${d\over d_\qt w} f(w) 
:= (f(\qt^\ha w)-f(\qt^{-\ha} w))/((\qt^\ha - \qt^{-\ha})w)$.
Here we use the identity
\be
\Exp{ \sum_{n>0}{1-t^n \over n} x^n }
- t
\Exp{ \sum_{n>0}{1-t^{-n} \over n} x^{-n} }
=
(1-t) \delta(x)
\ee
with the multiplicative $\delta$-function
$\delta(z):=\sum_{n\in\bZ} z^n$
satisfying
$\delta(z) f(z) = \delta(z) f(1)$.
Therefore the screening currents $S^a_\pm(z)$ 
commute with any {\qW} generators up to total difference.
Thus screening charges $\oint dz S^a_\pm(z)$ commute with any {\qW} generators
\be
[\oint dz S^a_\pm(z),W^b(w)]=0,
\qquad a,b = 1,2,\cdots,\nn -1.
\label{eq:SCandW}%
\ee
For a Laurent series $f(z):=\sum_{n\in\bZ} f_n z^n $ in $z$,
the integral ${\oint{dz\over \twoPiI z} f(z)}$ 
stands for the constant term in $f(z)$, i.e., 
\be
{\oint{dz\over \twoPiI z} }\sum_{n\in\bZ} f_n z^n 
:=
\CT{z}{ \sum_{n\in\bZ} f_n z^n }
:= f_0.
\ee
If $f$ is multivalued function,
we should choose an appropriate cycle 
or need to introduce a pseudo-constant to make it single-valued one.
(see \eqref{eq:qWDeltaSC}).%
\footnote{
One can replace the integral $\oint dz$ by any linear map which satisfies
$\oint dz {d\over d_{\qt} z} f(z) = 0 $ with $\qt =\qt_\pm$,
for example, by the Jackson integral, 
provided $f(0)=f(1)$.
}{ }


\subsection{Primary fields and degenerate operators}


For parameters $\tN$ and $\gamma$ with $\tN := t^\gamma$,
let us define the following vertex operators
\be
\VO^a_u(z) := 
\NP \Exp{ \sum_{n\neq0}
{ 
( \tN^{n\over 2} - \tN^{-{n\over 2}} ) 
{\Lam an} \over 
( q^{n\over 2} - q^{-{n\over 2}} ) 
( t^{n\over 2} - t^{-{n\over 2}} ) 
}
z^{-n} 
}\NP \,
e^{-{\gamma\rb}{\QLa a}} z^{-{\gamma\rb}{\Lam a0}}.
\label{eq:uqtPrimary}%
\ee
They satisfy 
\be
g^{a,L}_{\tN,p}\left({w\over z}\right) \Lambda_i(z) 
\VO^a_\tN(w) 
- 
\VO^a_\tN(w) 
\Lambda_i(z) g^{a,R}_{\tN,p}\left({z\over w}\right) 
=
(\tN^{-1} - 1) 
\sum_{b=1}^a \delta_{i,b}
\ddelta\left({w\over z \tN^{\ha}}\right) 
\NP \Lambda_i(z) \VO^a_\tN(w)\NP
,
\ee
where
$g^{a,L}_{\tN,p}(x)$ and 
$g^{a,R}_{\tN,p}(x)$
are inverse of the OPE factors,
\ba
g^{a,L}_{\tN,p}(x) 
&:=& 
{
\NP
\Lambda_j(z) 
\VO^a_\tN(w) 
\NP
\over 
\Lambda_j(z) 
\VO^a_\tN(w) 
}
=
\Exp{\sum_{n>0} { \tN^{-{n\over 2}} - \tN^{{n\over 2}}  \over n}
{ { \qpint a {p^n} } \over { \qpint \nn {p^n} } }
p^{ {n\over 2}( a_-\nn ) }x^n }
\tN^{ -{a\over \nn} },
\cr
g^{a,R}_{\tN,p}(x) 
&:=& 
{
\NP
\VO^a_\tN(w) 
\Lambda_j(z) 
\NP
\over 
\VO^a_\tN(w) 
\Lambda_j(z) 
}
=
\Exp{\sum_{n>0} { \tN^{{n\over 2}} - \tN^{-{n\over 2}}  \over n}
{ { \qpint a {p^n} } \over { \qpint \nn {p^n} } }
p^{ {n\over 2}( \nn - a ) }x^n }
\ea
for any $j>a$. 
Since $p^{D_z}$ commutes with $\VO^a_\tN(w)$, 
\ba
\left(p^{D_z} - g^{a,L}_{\tN,p}\left({w\over z}\right) \Lambda_i(z)\right)
\VO^a_\tN(w)
&-& 
\VO^a_\tN(w)
\left(p^{D_z} - \Lambda_i(z) g^{a,R}_{\tN,p}\left({z\over w}\right)\right)\cr
&=&
(1 - \tN^{-1}) 
\sum_{b=1}^a \delta_{i,b}
\ddelta\left({w\over z \tN^{\ha}} \right) 
\NP \Lambda_i(z) \VO^a_\tN(w)\NP.
\label{eq:pLambdaPhi}%
\ea
Remark that $\Lambda_{\nn-a+1}(z)$ and
$\VO^{\nn-a}_\tN(w)$ 
satisfy same relation
with replacing 
$p\leftrightarrow p^{-1}$ and
$\tN \leftrightarrow \tN^{-1}$.
By using \eqref{eq:pLambdaPhi} and $\omega_-$, we have
\\
{\bf Proposition.}~{\it 
The vertex operators $\VO^1_\tN(w)$ and $\VO^{\nn-1}_\tN(w)$ 
enjoy the following relations: 
\ba
&&\hspace{-5mm}
\NP
\left(p^{D_z} - g^{1,L}_{\tN,p}\left({w\over z       }\right) \Lambda_1(z       )\right)\cdots
 \left(p^{D_z} - g^{1,L}_{\tN,p}\left({w\over zp^{1-N}}\right) \Lambda_N(zp^{1-N})\right)
\NP
\VO^1_\tN(w)
\cr
&&
- \VO^1_\tN(w)
\NP
\left(p^{D_z} - \Lambda_1(z       ) g^{1,R}_{\tN,p}\left({z       \over w}\right)\right)\cdots
 \left(p^{D_z} - \Lambda_N(zp^{1-N}) g^{1,R}_{\tN,p}\left({zp^{1-N}\over w}\right)\right)
\NP
\cr
&& \hspace{5mm}
= 
(1 - \tN^{-1}) 
\ddelta\left({w\over z \tN^{\ha}} \right) 
\NP
\Lambda_1(z) \VO^1_\tN(w) 
\left(
p^{D_z} - \Lambda_2(zp^{-1})\right)\cdots\left(p^{D_z} - \Lambda_N(zp^{1-N})
\right)
\NP,~~~
\label{eq:PrimaryMiura}%
\ea
\ba
&&\hspace{-5mm}
\NP
\left(p^{D_z} - g^{1,L}_{1/\tN,1/p}\left({w\over z}\right) \Lambda_1(z )\right)
\cdots
\left(
p^{D_z} - g^{1,L}_{1/\tN,1/p}\left({w\over zp^{1-N}}\right) \Lambda_N(zp^{1-N})
\right)
\NP
\VO^{\nn-1}_\tN(w)
\cr
&&
- \VO^{\nn-1}_\tN(w)
\NP
\left(p^{D_z} - \Lambda_1(z)  g^{1,R}_{1/\tN,1/p} 
\left({z\over w}\right)\right)
\cdots
\left(
p^{D_z} - \Lambda_N(zp^{1-N}) g^{1,R}_{1/\tN,1/p}\left({zp^{1-N}\over w}
\right)\right)
\NP
\cr
&& \hspace{5mm}
= 
(1 - \tN) 
\ddelta\left({w\tN^{\ha}\over z p^{1-\nn}} \right) 
\NP
\left(p^{D_z} - \Lambda_1(z)\right)\cdots
\left(p^{D_z} - \Lambda_{N-1}(zp^{2-N})
\right)
\Lambda_{\nn}(zp^{1-\nn}) \VO^{\nn-1}_\tN(w) 
\NP.
\nonumber
\ea
}
Expanding \eqref{eq:PrimaryMiura}
gives the relation with the {\qW} generators $W^i(z)$.


When $\tN=t$ or $q^{-1}$,
let 
$\VO^a_+(z):=\VO^a_t(z)$ and
$\VO^a_-(z):=\VO^a_{q^{-1}}(z)$,
i.e.,
\be
\VO^a_\pm(z) := 
\NP \Exp{ \pm\sum_{n\neq0}
{ {\Lam an} \over \qt_\pm^{n\over 2} - \qt_\pm^{-{n\over 2}} }
z^{-n} 
}\NP \,
e^{\mp\rb^{\pm 1}{\QLa a}} z^{\mp\rb^{\pm 1}{\Lam a0}}
\label{eq:primaryPM}%
\ee
with $\qt_+ :=q$, $\qt_- :=t$. 
As a generalization of the the $(\ell+1,k+1)$ operators in the $\nn =2$ case 
\cite{rf:AKMOS},
we can define a $q$-deformation of the degenerate operators 
for $\ell,k\in\bZ_{\geq 0}$,
\ba
\VO^a_{\ell+1,k+1}(z) 
&:=& 
\NP
\prod_{i=1}^\ell \VO^a_+(q^{\ell+1-2i\over 2\ell}z)
\prod_{j=1}^ k   \VO^a_-(t^{ k  +1-2j\over 2 k  }z)
\NP
\cr
&=&
\NP \Exp{ \sum_{n\neq0}
\left(
{ 1\over q^{n\over 2\ell} - q^{-{n\over 2\ell}} } 
-
{ 1\over t^{n\over 2 k  } - t^{-{n\over 2 k  }} } 
\right)
{\Lam an} z^{-n}
}\NP \,
e^{{\alphalk \ell k}{\QLa a}} z^{{\alphalk \ell k} {\Lam a0}}
\label{eq:primaryellk}%
\ea
with
${\alphalk \ell k} := -\ell\rb + k/\rb$
and 
${1/( q^{n\over 2\ell} - q^{-{n\over 2\ell}}) } \vert_{\ell=0} := 0$.


\subsection{Boson Fock space}


Next we refer to the representation of the {\qW} algebra.
Let $\cF_\alpha$ be the boson Fock space 
generated by the highest weight state $|\alpha\rangle$ such that
${\al an} |0\rangle = 0$ for $n\geq0$ and 
$|\alpha\rangle := \exp\{\sum_{a=1}^{\nn -1}\alpha^a{\QLa a}\} |0\rangle$.
Note that 
$
{\al a0}|\alpha\rangle = \alpha^a|\alpha\rangle
$.
The dual module $\cF_\alpha^*$ is generated by $\langle \alpha|$ such that
$\langle 0| {\al a{-n}} = 0$ for $n\geq0$ and 
$\langle\alpha| := \langle 0| \exp\{-\sum_{a=1}^{\nn -1}\alpha^a{\QLa a}\}$.
The bilinear form $\cF_\alpha^*\otimes \cF_\alpha\rightarrow\bC$
is uniquely defined by $\langle 0|0\rangle=1$.


\subsection{Highest weight module of {\qW} algebra}


%
Let $|\lambda\rangle$ be the highest weight vector of the {\qW} algebra 
which satisfies
$W^a_n|\lambda\rangle=0$ for $n>0$ and $a=1,2,\cdots,\nn -1$ and 
$W^a_0|\lambda\rangle= \lambda^a |\lambda\rangle$ with $\lambda^a\in\bC$.
Let $M_\lambda$ be the Verma module over the {\qW} algebra
generated by $|\lambda\rangle$.
The dual module $M_\lambda^*$ is generated by $\langle\lambda |$ such that
$\langle \lambda |W^a_n=0$ for $n<0$ and 
$\langle\lambda |W^a_0= \lambda^a\langle\lambda |$.
The bilinear form $M_\lambda^*\otimes M_\lambda\rightarrow\bC$
is uniquely defined by $\langle\lambda |\lambda\rangle=1$.
A singular vector $|\chi\rangle\in M_\lambda$ is defined by 
$W^a_n|\chi\rangle=0$ for $n>0$ and 
$W^a_0|\chi\rangle= (\lambda^a+\nn^a) |\chi\rangle$ 
with $\nn^a\in\bC$.

%
%
 
The highest weight vector 
$|\alpha\rangle\in\cF_\alpha$ 
of the boson algebra is also that of the {\qW} algebra,
i.e.,
$W^a_n|\alpha \rangle=0$  for $n>0$ and $a=1,2,\cdots,\nn -1$. 
Note that
$
W^a_0|0 \rangle = [\nn ]_p^a |0\rangle
$
with 
$[\nn ]_p := ( p^{{\nn \over 2}} - p^{-{\nn \over 2}} )/( p^\ha - p^{-\ha} ) $.


\subsection{Singular vectors}


For a set of non-negative integers $s_a$ and $r_a\geq r_{a+1}\geq0$
with $a=1,\cdots,\nn -1$,
let 
\ba
         \pm \alpha_{r,s}^{\pm,a} 
&:=&(1+r_a-r_{a-1})\rb^{\pm1}-(1+s_a)\rb^{\mp1},\qquad
r_0 := 0,\cr
\pm\widetilde\alpha_{r,s}^{\pm,a} 
&:=&(1-r_a+r_{a+1})\rb^{\pm1}-(1+s_a)\rb^{\mp1},\qquad
r_\nn := 0.
\label{eq:weightalpha}%
\ea
The singular vectors $|\chi_{rs}^\pm\rangle\in\cF_{\alpha_{rs}^\pm}$ 
are realized by the screening currents as follows:
\ba
&&
\hspace{-5mm}
|\chi_{r,s}^\pm\rangle =
\oint\prod_{a=1}^{\nn -1}\prod_{j=1}^{r_a} {dz^a_j\overtwoPiI}\cdot
S^1_\pm(z^1_1)\cdots S^1_\pm(z^1_{r_1}) \cdots
S^{\nn -1}_\pm(z^{\nn -1}_1) \cdots S^{\nn -1}_\pm(z^{\nn -1}_{r_{\nn -1}}) 
|\widetilde\alpha_{r,s}^\pm\rangle
\cr
&=&
\oint\prod_{a=1}^{\nn -1} \prod_{j=1}^{r_a} {dz^a_j\over \twoPiI z^a_j}
(z^a_j)^{-s_a} (S^a_\pm(z^a_j))_-
\cdot
\qWDelta(z^a;\qt_\pm,\qt_\mp) 
\Pi\left(\overline{z^a},pz^{a+1};\qt_\pm,\qt_\mp\right) 
|\alpha_{r,s}^\pm\rangle
~~~
\label{eq:singular}%
\ea
with $z^\nn := 0$, $\overline z := 1/z$, $\qt_+ :=q$ and $\qt_- :=t$. 
Note that 
$\omega_-\omega_+ |\chi_{r,s}^+\rangle=|\chi_{r,s}^-\rangle$.
Here
\ba
\Pi(z,w) 
:=
\Pi(z,w;q,t) 
&:=& \prod_{i,j} 
\Exp{ \sum_{n>0}{1\over n}
{ t^{n\over 2}-t^{-{n\over 2}} \over q^{n\over 2}-q^{-{n\over 2}} } p^{-{n\over 2}}
z_i^n w_j^n }
\cr
&=& \prod_{i,j} \prod_{\ell\geq 0}
{1-q^\ell t z_i w_j \over 1-q^\ell z_i w_j },
\qquad |q|<1,
\\
\qWDelta(z) 
:=
\qWDelta(z;q,t) 
&:=&
\prod_{i<j}
\Exp{-\sum_{n>0}{1\over n}
{ t^{n\over 2}-t^{-{n\over 2}} \over q^{n\over 2}-q^{-{n\over 2}} }
( p^{n\over 2} + p^{-{n\over 2}} )
{z_j^n\over z_i^n}}
\cdot\prod_{i=1}^r z_i^{(r+1-2i)\beta}
~~~
\cr
&=&
\prod_{i<j}
(1- z_j/z_i)
\prod_{\ell\geq 0}
{1- q^\ell p  z_j/z_i \over 1-q^\ell t  z_j/z_i }
\cdot\prod_{i=1}^r z_i^{(r+1-2i)\beta},
\quad |q|<1
\label{eq:Delta}%
\ea
with $\beta := \log t/\log q$.
Note that 
$\qWDelta(cz) = \qWDelta(z)$.


When $\beta\notin\bZ$, 
$\qWDelta(z)$ becomes a multivalued function
but we can replace it with the following single valued one
\be
\widetilde{\qWDelta}(z) 
:=
\qWDelta(z) 
\prod_{i<j} F(z_j/z_i)
\label{eq:qWDeltaSC}%
\ee
with the pseudo-constant $F(x)=F(qx)$ defined by
\be
F(x):=
x^{\beta-1}
{ \vartheta_q(tx) \over \vartheta_q(qx) }.
\label{eq:pseudoF}%
\ee
Here
\be
\vartheta_q(x)
:=
\prod_{\ell\geq 0} (1-q^\ell x)(1-q^{\ell+1}/x)(1-q^{\ell+1})
\ee
is  the $\vartheta$-function with the multiplicative period $q$.


\Section{Quantum deformation of $\beta$-ensemble}
\label{secQGMM}%

Note that the singular vector in \eqref{eq:singular}
is naturally mapped to the Macdonald polynomial 
\cite{rf:Macdonald} 
defined in the appendix {\appMac} 
\cite{rf:Awata-Odake-Shiraishi}\cite{rf:AKOS}.
As a generalization of this map
one can define, under the strategy of \cite{rf:AMOS:Note},
a quantum deformation of the generalized matrix model,
i.e., $q$-deformed $\beta$-ensemble.


\subsection{Isomorphisms between bosons}


With a new parameters $p^{(a)}:=(p^{(a)}_1,p^{(a)}_2,\cdots)$
let us define the following vertex operator
\be
V_\nn
:=
\prod_{a=1}^{\nn -1} 
\Exp{\sum_{n>0}{{\Lam an}\over q^{n\over 2}-q^{-{n\over 2}}  } p^{(a)}_n }.
\label{eq:VnDef}%
\ee
Note that $[{\Lam an},{\Lam bm}]=0$ for $n,m>0$.
Then 
$\langle \alpha|V_\nn$ 
defines the isomorphism between 
the boson algebras 
$\langle {\h an}\rangle_{n\in\bZ}^{1\leq a < \nn }$
and 
$\langle p^{(a)}_n , \alpha^a,\deldel{ p^{(a)}_n } \rangle_{n\in\bN}^{1\leq a < \nn }$
by
\ba
{ t^{n\over 2}-t^{-{n\over 2}} \over n } p^{(a)}_n 
\langle \alpha|V_\nn
&=&
\langle \alpha|V_\nn
{\al a{-n}} ,
\cr
( q^{n\over 2}-q^{-{n\over 2}} )\deldel{ p^{(a)}_n } 
\langle \alpha|V_\nn
&=&
\langle \alpha|V_\nn
{\Lam an}
\label{eq:isoDef}%
\ea
for $n>0$ and 
$\alpha^a
\langle \alpha|V_\nn
=\langle \alpha|V_\nn
{\al a{0}}
$.
Since 
${\h in} 
= \sum_{b=1}^{\nn -1} {\Thalpha ib{p^n} } {\al bn}
= \sum_{b=1}^{\nn -1} {\ThLambda ib{p^n} } {\Lam bn}$,
we have
\ba
{ t^{n\over 2}-t^{-{n\over 2}} \over n } 
\sum_{b=1}^{\nn -1} {\Thalpha ib{p^{-n}} } p^{(b)}_n 
\langle \alpha|V_\nn
&=&
\langle \alpha|V_\nn
{\h i{-n}} ,
\cr
( q^{n\over 2}-q^{-{n\over 2}} )
\sum_{b=1}^{\nn -1} {\ThLambda ib{p^n} } \deldel{ p^{(b)}_n } 
\langle \alpha|V_\nn
&=&
\langle \alpha|V_\nn
{\h in}
\label{eq:isoH}%
\ea
for $n>0$ and
$
h^i
\langle \alpha|V_\nn
=
\langle \alpha|V_\nn
{\h i{0}}
$
with
$
h^i = 
\left[ 
\sum_{b=i}^{\nn -1} - \sum_{b=1}^{\nn -1}b/\nn 
\right] \alpha^b
$.


The vector
$| S_{r,s}^+\rangle := \prod_{a=1}^{\nn -1} \prod_{k=1}^{r_a} 
(S^a_+(z^a_k))_-\cdot | \alpha_{r,s}^+\rangle$
in \eqref{eq:singular}
also defines another linear map from  
$\langle {\h an}\rangle_{n\in\bN}^{1\leq a < \nn }$
to
$\langle  \sum_{k=1}^{r_a} (z^a_k)^n \rangle^{1\leq a < \nn}$
by
\be
{\Lam an}
| S_{r,s}^+\rangle
=
| S_{r,s}^+\rangle
{ t^{n\over 2}-t^{-{n\over 2}} \over n } \sum_{k=1}^{r_a} (z^a_k)^n,
\qquad
n>0.
\label{eq:LinearMap}%
\ee
Then
\be
{\h in}
| S_{r,s}^+\rangle
=
| S_{r,s}^+\rangle
{ t^{n\over 2}-t^{-{n\over 2}} \over n }  
\sum_{b=1}^{\nn -1} {\ThLambda ib{p^n} }
\sum_{k=1}^{r_b} (z^b_k)^n,
\qquad
n>0.
\ee


\subsection{$q$-deformed $\beta$-ensemble}



Let us define the following partition function
\\
{\bf Definition.}~{\it 
Let
$
Z_\nn
:=
Z_\nn\left(\{p^{(a)}\}_{a=1}^{\nn-1}\right)
:=
\langle \alpha_{r,s}^+|V_\nn |\chi_{r,s}^+\rangle
$.
}

\noindent
Then by
\eqref{eq:singular},
\eqref{eq:SCDef} and
\eqref{eq:isoDef},
we have
\ba
Z_\nn
&=&
\oint\prod_{a=1}^{\nn -1}\prod_{j=1}^{r_a} {dz^a_j\overtwoPiI}
\langle \alpha_{r,s}^+|V_\nn
S^1_+(z^1_1)\cdots S^1_+(z^1_{r_1}) \cdots
S^{\nn -1}_+(z^{\nn -1}_1) \cdots S^{\nn -1}_+(z^{\nn -1}_{r_{\nn -1}}) 
|\widetilde\alpha_{r,s}^+\rangle
\cr
&=& 
\oint\prod_{a=1}^{\nn -1}\prod_{j=1}^{r_a} {dz^a_j \over \twoPiI z^a_j}
(z^a_j)^{-s_a}
\Exp{\sum_{n>0}{1\over n}
{ t^{n\over 2}-t^{-{n\over 2}} \over q^{n\over 2}-q^{-{n\over 2}} }(z^a_j)^n p^{(a)}_n }
\cdot
\qWDelta(z^a)
\Pi\left(\overline{z^a},pz^{a+1}\right) 
\cr
&=& 
\oint\prod_{a=1}^{\nn -1} \prod_{j=1}^{r_a} {dz^a_j\overtwoPiI}
\cdot
\prod_{a=1}^{\nn -1}
\qWDelta(z^a)
e^{W(z^a,z^{a+1})}
,
\label{eq:qGMM}%
\ea
\be
W(z^a,z^{a+1})
:= 
\sum_{n>0}{1\over n}
{ t^{n\over 2}-t^{-{n\over 2}} \over q^{n\over 2}-q^{-{n\over 2}} }
\sum_{i=1}^{r_a}
\left\{
(z^a_i)^n p^{(a)}_n 
+
\sum_{j=1}^{r_{a+1}}
\left({ p^\ha z^{a+1}_j \over z^a_i }\right)^n 
\right\}
-
(s_a+1)
\sum_{i=1}^{r_a}
\log z^a_i
.
\label{eq:qSuperPotentioal}%
\ee
Here $z^\nn := 0$.
This $Z_\nn$ is regarded as a $q$-deformation of 
the partition function of the generalized matrix model \cite{rf:AMOS:Note},
i.e., $\beta$-ensemble.
One can define other type of partition functions 
by acting involutions 
\eqref{eq:Involqttq}, \eqref{eq:Involqqinv} and \eqref{eq:InvolMac}.
%
%


We can calculate this integral by using 
the Macdonald polynomials $P_\lambda(x)$ with the Young diagram $\lambda$,
their fusion coefficient $ f_{\lambda,\mu}^{\nu}$
and the inner products
$\IP **{q,t}$,
$\MacIP **r$ and 
$\qWIP **r$
defined in the appendix {\appMac}.
By  the Cauchy formula \eqref{eq:AppCauchy},
the Galilean boost \eqref{eq:GalileanBoost}
and \eqref{eq:FusionCoeff}
we have
\ba
Z_\nn
&=& 
\oint\prod_{a=1}^{\nn -1} \prod_{j=1}^{r_a} {dz^a_j\over \twoPiI z^a_j}
\cdot
P_{(s_a^{r_a})}(\bar z^a)
 \sum_{\lambda_a} 
 { 
P_{\lambda_a}(z^a) P_{\lambda_a}({\xp{p^a}})
 \over 
 \PIP{\lambda_a}
 }
\qWDelta(z^a)
 \sum_{\mu_a} 
  {
 P_{\mu_a}(\bar z^a)P_{\mu_a}(p z^{a+1})
 \over 
 \PIP{\mu_a}
 } 
\cr
&=&
\oint\prod_{a=1}^{\nn -1} \prod_{j=1}^{r_a} {dz^a_j\over \twoPiI z^a_j}
\cdot
 \sum_{\lambda_a,\mu_a,\nu_a} 
 f_{\mu_{a-1},\lambda_a}^{\nu_a}
  P_{\nu_a}(z^a)
 {
P_{\lambda_a}({\xp{p^a}})
 \over 
 \PIP{\lambda_a}
 } 
\qWDelta(z^a)
p^{|\mu_a|} 
 {
 P_{ \mu_a+(s_a^{r_a}) }(\bar z^a)
 \over 
 \PIP{\mu_a}
 }
 \ea
with 
$\mu_0 = \mu_{\nn-1} := (0)$.
Here $\lambda_a$, $\mu_a$ and $\nu_a$ are Young diagrams
such that 
$\lambda_{a,i} \geq \lambda_{a,i+1}$, and so on.
$P_{\lambda}({\xp p})$
denotes the Macdonald function in power sums $p:=(p_1,p_2,\cdots)$.
By the orthogonality with respect to the inner product $\qWIP **r$ 
in \eqref{eq:qWInnerProduct}, we obtain
\\
{\bf Proposition.}~{\it 
\be
Z_\nn 
=
\prod_{a=1}^{\nn -1}
 \sum_{\lambda_a,\mu_a} 
 f_{\mu_{a-1},\lambda_a}^{\mu_a+(s_a^{r_a})}
  P_{\mu_a+(s_a^{r_a})}(z^a)
 {
P_{\lambda_a}({\xp{p^a}})
 \over 
 \PIP{\lambda_a}
 } 
p^{|\mu_a|} 
 {
r_a!\qWPIP{ \mu_a+(s_a^{r_a}) }{r_a}
 \over 
 \PIP{\mu_a}
 }
\label{eq:ZbyMac}%
\ee
with
$\PIPbk{ 0 } := 1$.
}

For any $\lambda$ with $\lambda_1\leq s$,
let $\widehat\lambda$ be its complements with respect to $(s^\mm)$,
i.e., $\widehat\lambda_i = s-\lambda_{\mm-i+1}$. 
The fusion coefficient $f_{\lambda,\mu}^\nu$ 
defined in \eqref{eq:FusionCoeff} satisfies
$
f_{\lambda,(0)}^{\nu}
=
\delta_{\lambda,\nu}
$
and
$
f_{\lambda,\mu }^{(s^r)}
=
\delta_{\mu,\widehat\lambda}
f_{\lambda,\widehat\lambda }^{(s^r)}
$.
Since 
$\mu_0 = \mu_{\nn-1} = (0)$,
we have 
$\lambda_1 = \mu_1+(s_1^{r_1}) $ and 
$\lambda_{\nn-1} = \widehat\mu_{\nn-2}$
with respect to $(s_{\nn-1}^{r_{\nn-1}})$.
Therefore \eqref{eq:ZbyMac} is summed over 
$(\nn-2)+(\nn-3)$ Young diagrams for $\nn\geq 3$.


For any function $\cO$ in $z^a_j$'s,
the correlation function with respect to $\cO$ 
is defined by 
\ba
\CF{ \cO }
:=
{1\over Z_\nn}
\oint\prod_{a=1}^{\nn -1} \prod_{j=1}^{r_a} {dz^a_j\overtwoPiI}
\cdot
\cO
\prod_{a=1}^{\nn -1}
\qWDelta(z^a)
e^{W(z^a,z^{a+1})}
.
\ea
%
%
The effective action $S_{\rm eff}$ defined by 
$Z_\nn
=:
\oint\prod_{a=1}^{\nn -1}\prod_{j=1}^{r_a} {dz^a_j\overtwoPiI}
\cdot
e^{S_{\rm eff}}$
is now
\be
S_{\rm eff}
=
\sum_{a=1}^{\nn-1}
W(z^a,z^{a+1})
-
\sum_{n>0}{ {\qpint 2{p^n}}\over n }
{ t^{n\over 2}-t^{-{n\over 2}} \over q^{n\over 2}-q^{-{n\over 2}} } 
\sum_{a=1}^{\nn -1}
\sum_{i<j}
\left({ z^{a}_j \over z^{a}_i }\right)^n
+
\beta
\sum_{a=1}^{\nn -1}
\sum_{i=1}^{r_a}
(r_a+1-2i)
\log z^{a}_i
.
\label{eq:Seff}%
\ee
The saddle point condition is
${\partial S_{\rm eff} \over \partial z^a_k } = 0 $
with
\ba
z^a_k 
{\partial S_{\rm eff} \over \partial z^a_k }
&=&
\sum_{n>0}{ t^{n\over 2}-t^{-{n\over 2}} \over q^{n\over 2}-q^{-{n\over 2}} } 
\left\{
(z^a_k)^n 
p^{(a)}_n 
-
{\qpint 2{p^n}}
\left(
\sum_{i<k}
\left({ z^{a}_k \over z^{a}_i }\right)^n
-
\sum_{j>k}
\left({ z^{a}_j \over z^{a}_k }\right)^n
\right)
\right.
\cr
&+&
\left.
\sum_{i=1}^{r_{a-1}}
\left({ p^\ha z^{a}_k \over z^{a-1}_i }\right)^n
-
\sum_{j=1}^{r_{a+1}}
\left({ p^\ha z^{a+1}_j \over z^{a}_k }\right)^n
\right\} 
+
((r_a+1-2k)\beta - s_a-1)
.
\label{eq:saddlePt}%
\ea


\subsection{ {\qW} constraint}


Next let us define  $\cLa_i(z)$ and $ \cW^i(z)$ as follows, 
which are the power sum realization of fundamental vertices $\Lambda_i(z)$ and 
{\qW} generators $W^i(z)$, respectively:
\ba
\cLa_i(z)
&:=&
\Exp{\sum_{n>0}{ t^{n\over 2}-t^{-{n\over 2}} \over n }z^n 
\sum_{b=1}^{\nn -1} {\Thalpha ib{p^{-n}} } p^{(b)}_n 
}
\cr
&\times&
\Exp{\sum_{n>0}( q^{n\over 2}-q^{-{n\over 2}} ) z^{-n} 
\sum_{b=1}^{\nn -1} {\ThLambda ib{p^n} } \deldel{ p^{(b)}_n } 
}
q^{\rb{\h i{r,s}}} p^{{\nn +1\over 2}-i},
\ea
\be
\sum_{i=0}^\nn (-1)^i \cW^i(zp^{1-i\over 2}) p^{(\nn -i)D_z}
:=
\NP \left(p^{D_z} - \cLa_1(z)\right) 
\left(p^{D_z} - \cLa_2(zp^{-1})\right) 
\cdots
\left(p^{D_z} - \cLa_\nn (zp^{1-\nn })\right)\NP
\label{eq:ConstraintMiura}%
\ee
and 
$\cW^i(z)=:\sum_{n\in\bZ} \cW^i_n z^{-n}$.
Here $\NP *\NP $ stands for the normal ordering such that
the differential operators $\deldel {p_n^{(a)}}$ 
are in the right.
Then by the isomorphism \eqref{eq:isoH},
\be
\cLa_i(z)
\langle \alpha_{r,s}|V_\nn
=
\langle \alpha_{r,s}|V_\nn
\Lambda_i(z),
\qquad
\cW^i(z)
\langle \alpha_{r,s}|V_\nn
=
\langle \alpha_{r,s}|V_\nn
W^i(z).
\ee

Therefore the highest weight condition for the singular vector
$W^a_n|\chi\rangle=0$ for $n>0$ is equivalent to the following {\qW} constraint: 
\\
{\bf Theorem.}
\be
\cW^a_n Z_\nn = 0,
\qquad
n>0.
\label{eq:qWconstraint}%
\ee


\subsection{Loop equation and quantum spectral curve}


Let us define  $\tcLa_i(z)$ and $ \tcW^i(z)$ as follows, 
which correspond to fundamental vertices $\Lambda_i(z)$ and 
{\qW} generators $W^i(z)$, respectively:
\ba 
\tcLa_i(z) 
&:=& 
\Exp{\sum_{n>0}{ t^{n\over 2}-t^{-{n\over 2}} \over n }z^n
\sum_{b=1}^{\nn -1} {\Thalpha ib{p^{-n}} } p^{(b)}_n 
} 
\cr 
&\times& 
\Exp{\sum_{n>0}
{ t^{n\over 2}-t^{-{n\over 2}} \over n } z^{-n} 
\sum_{b=1}^{\nn -1} {\ThLambda ib{p^n} } 
\sum_{k=1}^{r_b} (z_k^{b})^n 
} 
q^{\rb{\h i{r,s}}} p^{{\nn +1\over 2}-i}, 
\ea 
\be
\sum_{i=0}^\nn (-1)^i \tcW^i(zp^{1-i\over 2}) p^{(\nn -i)D_z}
:=
\left(p^{D_z} - \tcLa_1(z)\right)
\left(p^{D_z} - \tcLa_2(zp^{-1})\right) \cdots 
\left(p^{D_z} - \tcLa_\nn (zp^{1-\nn })\right) 
\label{eq:LoopMiura}%
\ee 
and
$\tcW^i(z)=:\sum_{n\in\bZ} \tcW^i_n z^{-n}$.  
Then by linear maps \eqref{eq:isoH} and \eqref{eq:LinearMap},
we have 
\ba
\langle \alpha_{r,s}^+|V_\nn\Lambda_i(z)| S_{r,s}^+\rangle
&=&
\langle \alpha_{r,s}^+|V_\nn | S_{r,s}^+\rangle
\tcLa_i(z),
\cr
\langle \alpha_{r,s}^+|V_\nn W^i(z)| S_{r,s}^+\rangle
&=&
\langle \alpha_{r,s}^+|V_\nn | S_{r,s}^+\rangle
\tcW^i(z).
\ea
Hence
\be
{1\over Z_\nn}
\langle \alpha_{r,s}^+|V_\nn W^i(z)|\chi_{r,s}^+\rangle 
=
\CF{ \tcW^i(z) }. 
\ee
Therefore the highest weight condition for the singular vector
$W^a_n|\chi\rangle=0$ for $n>0$ is equivalent to the following loop equation: 
\\
{\bf Theorem.}
\be
\CF{ \tcW^a_n }= 0,
\qquad n>0.
\label{eq:LoopEq}%
\ee


Note that, although the variables $z^{(a)}_j$, $z$ and $x^{(a)}_j$
are all formal parameters,
one can treat them as complex parameters with 
\be
\infty 
> |x^{(a)}_j|^{-1} > |z| > |z^1_1| 
> \cdots > |z^1_{r_1}| 
> \cdots > |z^{\nn -1}_1|
> \cdots > |z^{\nn -1}_{r_{\nn -1}}| 
> 0. 
\label{eq:ordering}%
\ee
Here 
$| z^a_i | > | z^a_{i+1} |$ and 
$| z^a_i | > | z^{a+1}_j |$.

The quantum spectral curve should be
\be
\CF{
 \left(p^{D_z} - \tcLa_1(z)\right) 
\left(p^{D_z} - \tcLa_2(zp^{-1})\right) 
\cdots
\left(p^{D_z} - \tcLa_\nn (zp^{1-\nn })\right)
}
=0
\label{eq:SpectralCurve}%
\ee
which regularity in $z$ is guaranteed by the loop equation
\eqref{eq:LoopEq}.


\subsection{Large $r_a$ case}


Let 
$(q,t)
=:
(e^{R \epsilon_2},e^{-R \epsilon_1})
=:
(e^{\gs R},e^{\gs \beta R})
$
with the radius 
$R$ of the 5th dimensional circle $S^1$.
Let us rescale the variables as
$\tp^{(a)}_n := \gs p^{(a)}_n$, 
$\tilk{r_a} := \gs r_a$
and 
$\tilk{s_a} := \gs s_a$. 
Then
\ba 
\tcLa_i(z) 
&=& 
\Exp{\sum_{n>0}
{ t^{n\over 2}-t^{-{n\over 2}} \over n \gs } z^n
\sum_{b=1}^{\nn -1} {\Thalpha ib{p^{-n}} } \tp^{(b)}_n 
} 
\cr 
&\times& 
\Exp{\sum_{n>0}
{ t^{n\over 2}-t^{-{n\over 2}} \over n \gs } z^{-n} 
\sum_{b=1}^{\nn -1} {\ThLambda ib{p^n} } 
\int dw w^n  \tilr_b \rho_b(w)
} 
q^{\rb{\h i{r,s}}} p^{{\nn +1\over 2}-i}
\ea 
with
$\rho_a(w)
:= 
{1\over r_a}\sum_{k=1}^{r_a} \delta(w-z^a_k)
$.
Note that by \eqref{eq:ordering},
\ba
\sum_{i<k}
\left({ z^{a}_k \over z^{a}_i }\right)^n
-
\sum_{j>k}
\left({ z^{a}_j \over z^{a}_k }\right)^n
&=&
r_a
\left[
\int_{|w|>|z^{a}_k|} dw 
\left({ z^{a}_k \over w }\right)^n
-
\int_{|w|<|z^{a}_k|} dw 
\left({ w \over z^{a}_k }\right)^n
\right]\rho_a(w),
\cr
\sum_{n>0}
\left\{
\sum_{i<k}
\left({ z^{a}_k \over z^{a}_i }\right)^n
-
\sum_{j>k}
\left({ z^{a}_j \over z^{a}_k }\right)^n
\right\}
&=&
r_a
\left[
\int_{|w|>|z^{a}_k|} dw 
+
\int_{|w|<|z^{a}_k|} dw 
\right]
{z^a_k\rho_a(w) \over w-z^a_k }
+r_a-k 
\cr
&=&
r_a 
\dashint dw 
{z^a_k\rho_a(w) \over w-z^a_k }
+r_a-k .
\ea


Under the limit 
$\gs \rightarrow 0$ and 
$r_a$, $s_a$, 
$k\rightarrow \infty$
with fixed 
$\tilk{r_a} := \gs r_a$,
$\tilk{s_a} := \gs s_a$
and 
$\tilk{k} = \gs k$,
the saddle point condition becomes
\ba
0&=&
\beta
\sum_{n>0}
\left\{
z^n 
\tp^{(a)}_n 
-
2 \tilr_a
\left[
\int_{|w|>|z|} dw 
\left({ z \over w }\right)^n
-
\int_{|w|<|z|} dw 
\left({ w \over z }\right)^n
\right]\rho_a(w) 
\right.
\cr
&&\hskip36pt
+
\left.
\tilr_{a-1}
\int dw \rho_{a-1}(w) 
\left({ z \over w }\right)^n
-
\tilr_{a+1} 
\int dw \rho_{a+1}(w) 
\left({ w \over z}\right)^n
\right\} 
+
\beta(
\tilr_a 
- 2\tilk{k}
)
-\tilk{s_a}
\cr
&=&
\beta
\left[
\tilr_{a-1}
\int dw\rho_{a-1}(w)
-
2\tilr_{a}
\dashint dw\rho_{a}(w)
+
\tilr_{a+1}
\int dw\rho_{a+1}(w)
\right]
{z\over w-z}
\cr
&&\hskip36pt
+\beta
\sum_{n>0} z^n \tp^{(a)}_n 
+
\beta(
\tilr_{a+1}
-\tilr_{a}
)
-\tilk{s_a}
\label{eq:saddlePtLarge}%
\ea
with 
$z:=\lim_{k\rightarrow\infty} z_k$ and
$\rho_a(w)
:= 
\lim_{r_a\rightarrow\infty}{1\over r_a}\sum_{j=1}^{r_a} \delta(w-z^a_j)
$.
%
Under this limit,
the sift operator $p^{D_z}$ tends to a commutative variable, say $\tz$,
and the spectral curve reduces to 
\be 
\prod_{i=1}^\nn \left(\tz - \tcLa_i(z)\right)
=
0,
\label{eq:LargeRSpectralCurve}%
\ee
$$
\tcLa_i(z) 
= 
\Exp{\beta R \left\{
\sum_{n>0} z^n
  \left[
\sum_{a=i}^{\nn -1}  - \sum_{a=1}^{\nn -1} {a\over \nn } 
\right]
\tp^{(a)}_n 
+ 
\int {dw w\over z-w } \left( \tilr_i \rho_i(w) - \tilr_{i-1} \rho_{i-1}(w)\right)
\right\}
} 
$$
with the solution $\rho_a(w)$ of \eqref{eq:saddlePtLarge}.
Note that the parameter $\beta$ appears only in the combinations
$\beta\tilr_a$ and $\beta\tp^{(a)}_n$.

\newpage
\subsection{$q$-deformed Liouville correlation function}


In the relation with the Macdonald polynomial in \cite{rf:AKOS}, 
the parameter $p_n$ is mapped to the power sum
$p_n = \sum_{i=1}^\mm x_i^n$.
On the other hand, the principal specialization 
$x_i = t^{i-1}$, i.e., 
$p_n = {(1-t^{\mm n})/(1-t^n)}$,
has a natural generalization 
$p_n = {(1-u^n)/(1-t^n)}$ with $u\in\bC$.
By these mapping and specialization, 
we can identify our partition function 
with a $q$-deformed Liouville correlation function.

With parameters $x^{(a)}_j$, $y^{(a)}$ and $\tN^{(a)}$ 
for $a=1,2,\cdots,\nn-1$ and $j=1,2,\cdots,M_a$,
let us consider the case that
\be
p^{(a)}_n
= 
\sum_{j=1}^{M_a} (x^{(a)}_j)^n 
+
{ 
(\tN^{(a)})^{{n\over 2}}-(\tN^{(a)})^{-{n\over 2}}
\over 
t^{n\over 2} - t^{-{n\over 2}} 
}
(y^{(a)})^n 
.
\ee
Then
\be
V_\nn
=
\prod_{a=1}^{\nn -1} 
\Exp{
\sum_{n>0}
{\Lam an}
\left\{
\sum_{j=1}^{M_a} 
{
(x^{(a)}_j)^n 
\over 
q^{n\over 2} - q^{-{n\over 2}}
}
+
{
(\tN^{(a)})^{{n\over 2}}-(\tN^{(a)})^{-{n\over 2}}
\over 
( q^{n\over 2} - q^{-{n\over 2}} ) 
( t^{n\over 2} - t^{-{n\over 2}} )
}
(y^{(a)})^n 
\right\}
}
\ee
is the positive mode part 
$\left(\VO^a_{\tN^{(a)}}(1/y^{(a)})\right)_+$ 
and
$\left(\VO^a_+(1/x^{(a)}_j)\right)_+$ 
of the primary fields  
$\VO^a_{\tN^{(a)}}(1/y^{(a)})$ 
in \eqref{eq:uqtPrimary}
and the $(2,1)$ operator
$\VO^a_+(1/x^{(a)}_j)$ 
in \eqref{eq:primaryPM}, respectively. 
Thus the partition function 
\ba
Z_\nn
&=&
\oint\prod_{a=1}^{\nn -1}\prod_{j=1}^{r_a} {dz^a_j\overtwoPiI}
\cdot
\langle \alpha_{r,s}^+|
\prod_{a=1}^{\nn-1}
\left(\VO^a_{\tN^{(a)}}(1/y^{(a)})\right)_+
\prod_{j=1}^{M_a}
\left(\VO^a_+(1/x^{(a)}_j)\right)_+
\cr
&&\hskip90pt\times
S^1_+(z^1_1)\cdots S^1_+(z^1_{r_1}) \cdots
S^{\nn -1}_+(z^{\nn -1}_1) \cdots S^{\nn -1}_+(z^{\nn -1}_{r_{\nn -1}}) 
|\widetilde\alpha_{r,s}^+\rangle
~~~
\ea
is nothing but 
the integral part of the $q$-deformed Liouville correlation function of them,
\ba
&&
\oint\prod_{a=1}^{\nn -1}\prod_{j=1}^{r_a} {dz^a_j\overtwoPiI}
\cdot
\langle \alpha_{r,s}^+|
\prod_{a=1}^{\nn-1}
\VO^a_{\tN^{(a)}}(1/y^{(a)})
\prod_{a=1}^{\nn-1}
\prod_{j=1}^{M_a}
\VO^a_+(1/x^{(a)}_j)
\cr
&&\hskip90pt\times
S^1_+(z^1_1)\cdots S^1_+(z^1_{r_1}) \cdots
S^{\nn -1}_+(z^{\nn -1}_1) \cdots S^{\nn -1}_+(z^{\nn -1}_{r_{\nn -1}}) 
|\widetilde\alpha_{r,s}^+\rangle
\cr
&&\hskip36pt
=
f(x,y) Z_\nn
.~~~
\ea
To recover the hole correlation function,
one just need to multiply the OPE factor $f(x,y)$
coming from the negative mode part.
Note that if $\tN^{(a)}=0$ and $M_a<\infty$ 
then $\{p^{(a)}_n\}_{n\in\bN}$ is linearly dependent
and thus {\qW} constraint \eqref{eq:qWconstraint} should be modified
but the loop equation \eqref{eq:LoopEq}
and the spectral curve \eqref{eq:SpectralCurve} are unchanged.

\Section{$\nn =2$ case} 
\label{sec\nn =2}%

Here we give an example when $\nn =2$, i.e., the $q$-deformed Virasoro case.
%
%
The fundamental bosons are
\be
[{\h {1}n},{\h {1}m}] = 
{1\over n}
{ (q^{{n\over 2}}-q^{-{n\over 2}})(t^{{n\over 2}}-t^{-{n\over 2}})
\over  p^{n\over 2} + p^{-{n\over 2}} }
	\ddelta_{n+m,0},
\qquad
[{\h {1}n},{\Qh {1}}] = \ha\ddelta_{n,0},
\ee
${\h 2n} := -p^{-n} {\h {1}n}$ and 
${\Qh 2} := - {\Qh {1}}$.
The root type and the weight type bosons are
${\al {1}n} := (1+p^{-n}) {\h {1}n}$,
${\Qal {1}} := 2{\Qh {1}}$,
${\Lam {1}n} :=  {\h {1}n} p^{-{n\over 2}}$ and 
${\QLa {1}} := {\Qh {1}}$.
Note that 
${\Thalpha 11p} = 1/(1+p^{-1})$,
${\ThLambda 11p} = p^\ha$ and 
${\TalphaLambda 11p} = {\qpint 2p}$.


The $q$-Virasoro generator, the screening currents and 
the vertex operators are now
\ba
W^{1}(z) &= & 
\NP \Exp{ \sum_{n\neq 0}{\h {1}n} z^{-n} }\NP \,
	q^{\rb{\h {1}0}} p^{\ha}
+ 
\NP \Exp{-\sum_{n\neq 0}{\h {1}n} p^{-n} z^{-n} }\NP \,
	q^{-\rb{\h {1}0}} p^{-\ha},
\cr
S^{1}_\pm(z) &=& 
\NP \Exp{ \mp\sum_{n\neq0}
{1+p^{-n}\over \qt_\pm^{n\over 2} - \qt_\pm^{-{n\over 2}} } {\h {1}n} z^{-n} }\NP \,
e^{\pm 2\rb^{\pm1}{\Qh {1}}} z^{\pm 2\rb^{\pm1}{\h {1}0}},
\qquad
\qt_+ = q,
\quad
\qt_- = t, 
\cr
\VO^1_\pm(z) 
&=& 
\NP \Exp{ \pm\sum_{n\neq0}
{ {\h {1}n} \over \qt_\pm^{n\over 2} - \qt_\pm^{-{n\over 2}} }
p^{-{n\over 2}} z^{-n} 
}\NP \,
e^{\mp\rb^{\pm 1}{\Qh {1}}} z^{\mp\rb^{\pm 1}{\h {1}0}},
\cr
\VO^1_u(z) 
&=& 
\NP \Exp{ \sum_{n\neq0}
{ 
( \tN^{n\over 2} - \tN^{-{n\over 2}} ) 
{\h 1n} \over 
( q^{n\over 2} - q^{-{n\over 2}} ) 
( t^{n\over 2} - t^{-{n\over 2}} ) 
}
p^{-{n\over 2}}z^{-n} 
}\NP \,
e^{-{\gamma\rb}{\Qh 1}} z^{-{\gamma\rb}{\h 10}},
\cr
V_2  
&=&
\Exp{\sum_{n>0}{{\h {1}n}\over q^{n\over 2}-q^{-{n\over 2}}  } 
p^{-{n\over 2}} p_n }.
\ea
For non-negative integers $s$ and $r\geq0$,
the singular vectors $|\chi_{rs}\rangle\in\cF_{\alpha_{rs}}$ are 
\ba
|\chi_{r,s}^+\rangle 
&=&\oint\prod_{j=1}^{r} {dz_j\overtwoPiI}\cdot
S^1_+(z_1)\cdots S^1_+(z_{r}) |\widetilde\alpha_{r,s}^+\rangle\cr
&=&
\oint\prod_{j=1}^{r} {dz_j\over \twoPiI z_j}
z_j^{-s} (S^{1}_+(z_j))_-
\cdot
\qWDelta(z) 
|\alpha_{r,s}^+\rangle
\ea
with $\alpha_{r,s}^{+,1} :=\rb(1+r)\rbi(1+s)$.
Here $\qWDelta(z)$ is same as \eqref{eq:Delta}.
%
%
The partition function $Z_2 $ is now
\ba
 Z_2(p) 
&=&
\oint \prod_{j=1}^{r} {dz_j\over \twoPiI z_j}
z_j^{-s}
\Exp{\sum_{n>0}{1\over n}
{ t^{n\over 2}-t^{-{n\over 2}} \over q^{n\over 2}-q^{-{n\over 2}} }
z_j^n p_n }
\cdot
\qWDelta(z)
\cr
&=&
p^{ rs \over 2 }
{ 
r!\qWPIPbk{ s^r }r
\over 
\PIPbk{ s^r }
}
P_{(s^r)}({\xp p}).
\label{eq:Ztwo}%
\ea


If we divide the power sum $p_n$ as $p_n =: p^{(1)}_n + p^{(2)}_n$
then by \eqref{eq:FusionFormula} and \eqref{eq:FusionComplement}
we have
\ba
Z_2(p^{(1)} + p^{(2)})
&=&
p^{ rs \over 2 }
\sum_{\lambda\subset (s^r)}
{ 
r!\qWPIPbk{ s^r }r
\over 
\PIP{ \lambda }
\PIP{ \widehat\lambda }
}
f_{\lambda,\widehat\lambda }^{(s^r)}
P_\lambda({ \xp{p^{(1)}} })
P_{\widehat\lambda} ({ \xp{p^{(2)}} })
\cr
&=&
p^{ rs \over 2 }
{
r!\qWPIPbk{ s^r }r
\over 
\MacPIPbk{ s^r }r
}
\sum_{\lambda\subset (s^r)}
{ 
\MacPIP{ \lambda }r
\over 
\PIP{ \lambda }
\PIP{ \widehat\lambda }
}
P_\lambda({ \xp{p^{(1)}} })
P_{\widehat\lambda} ({ \xp{p^{(2)}} }).
\ea
Here 
$\lambda\subset (s^r)$ means
$\lambda_1\leq s$ and $\ell(\lambda)\leq r$.


Let us consider the case that
$p^{(2)}_n :=
{ (1-u^n)y^n/(1-t^n) }$
and denote its function by
$f({\xp{ {1-u \over 1-t }y }})$.
Then we have
\be
{
P_{\widehat\lambda}\left({\xp{ {1-u \over 1-t }y }}\right)
\over 
\PIP{ \widehat\lambda }
y^{|\widehat\lambda|}
}
=
{
P_{(s^r)}\left({\xp{ {1-u \over 1-t }y }}\right)
\over 
\PIPbk{ s^r }
y^{rs} 
}
\prod_{(i,j)\in\lambda}
{
(1-q^j t^{r-i})
(1-q^{s-j+1}t^{i-1})
\over 
(t^{r-i}-uq^{s-j})
(1-q^{\lambda_i-j+1}t^{\lambda\dualvee_j-i})
},
\label{eq:ParaSpecialComplement}%
\ee
which is proved in the appendix {\appProof}.
Next, \eqref{eq:AnotherInnerProductValue} shows that
\be
{
\PIPbk{ s^r }
\over
\MacPIPbk{ s^r }r
}
{
\MacPIP{ \lambda }r
\over
\PIP{ \lambda }
}
=
\prod_{(i,j)\in (s^r)}
{ 1-q^j t^{r-i} \over 1-q^{j-1}t^{r-i+1} }
\prod_{(i,j)\in \lambda }
{ 1-q^{j-1}t^{r-i+1} \over 1-q^j t^{r-i} }.
\ee
{}From the above two equations, 
\be
{
1
\over
\MacPIPbk{ s^r }r
}
{
\MacPIP{ \lambda }r
\over
\PIP{ \lambda }
}
{
P_{\widehat\lambda}\left({\xp{ {1-u \over 1-t }y }}\right)
\over 
\PIP{ \widehat\lambda }
y^{|\widehat\lambda|}
}
=
{
P_{(s^r)}\left({\xp{ {1-u \over 1-t }y }}\right)
\over 
\PIPbk{ s^r }
y^{rs}
}
\prod_{(i,j)\in\lambda}
{
( 1-q^{j-1} t^{r-i+1} )
( 1-q^{s-j+1} t^{i-1} )
\over 
( t^{r-i}-u q^{s-j})
( 1-q^{\lambda_i-j+1}t^{\lambda\dualvee_j-i})
},
\ee
hence we obtain
\ba
&&\hskip-60pt
{ 
\PIPbk{ s^r }
\over
r!\qWPIPbk{ s^r }r
}
Z_2\left(p^{(1)}+{1-u\over 1-t}y\right)
=
p^{rs\over 2}
P_{(s^r)}\left({\xp{ {1-u \over 1-t }y }}\right)
\cr
&\times&
\sum_{\lambda\subset (s^r) }
y^{-|\lambda|}
P_\lambda({ \xp{p^{(1)}} })
\prod_{(i,j)\in\lambda}
{q\over u}
{
( t^{i-1}-q^{j-1} t^r )
( t^{i-1}-q^{j-1-s} )
\over 
( t^{i-1}-t^{r-1}q^{j-s}/u)
( 1-q^{\lambda_i-j+1}t^{\lambda\dualvee_j-i})
}.
\ea


The involution 
$\omega_{q,t}$ is defined in \eqref{eq:InvolMac}
as
$
\omega_{q,t} (p_n) 
=
(-1)^{n-1}p_n{(1-q^n)/(1-t^n)} 
$.
If we act $\omega_{q,t}$ 
on the variables $p_n$ in \eqref{eq:Ztwo}
and  denote it as 
$\omega_{q,t} Z_2(p^{(0)})
:=
\omega_{q,t} Z_2(p)\vert_{p=p^{(0)}}$,
then we get another type of formula
\ba
{ 1\over {r!\qWPIPbk{ s^r }r} }
\omega_{q,t}
Z_2(p)
&=&
{ 1\over {r!\qWPIPbk{ s^r }r} }
\oint \prod_{j=1}^{r} {dz_j\over \twoPiI z_j}
z_j^{-s}
\Exp{\sum_{n>0}{(-1)^{n-1}\over n}
v^n z_j^n p_n }
\cdot
\qWDelta(z)
\cr
&=&
p^{ rs \over 2 }
P_{(r^s)}({\xp p};t,q)
\cr
&=&
{ 
\PIPbk{ s^r }
\over
r!\qWPIPbk{ s^r }r
}
Z_2(p)
\vert_{ q\leftrightarrow t \atop r\leftrightarrow s}
\ea
which is just \eqref{eq:Ztwo} 
with the replacement $q\leftrightarrow t$ and $r\leftrightarrow s$.

Therefore when
$p^{(1)}_n := \sum_{i=1}^M x_i^n$
we obtain
\\
{\bf Proposition.}{\it~~ 
The partition function $Z_2(p)$ substituting 
$
p_n=\sum_i x_i^n
+{1-u^n\over 1-t^n}y^n$
and 
${ 1-t^n \over 1-q^n} p_n
=(-1)^{n-1}
(\sum_i x_i^n
+{1-u^n\over 1-t^n}y^n)$
are
\ba
{ 
Z_2 \left(\sum_i x_i+{1-u\over 1-t}y\right)
\over 
Z_2\left({1-u\over 1-t}y\right)
}
=
\qtHyperGeo 21qt{q^{-s},t^r}{q^{1-s}t^{r-1}/u}{{qx\over uy}},
\label{eq:qGMMVir}%
\\
\omega_{q,t}
{ 
Z_2 \left(\sum_i x_i+{1-u\over 1-t}y\right)
\over 
Z_2\left({1-u\over 1-t}y\right)
}
=
\qtHyperGeo 21tq{t^{-r},q^s}{t^{1-r}q^{s-1}/u}{{tx\over uy}}.
\label{eq:qGMMdualVir}%
\ea
Here
$\qtHyperGeo 21qt{a,b}cx$
is the multivariate $q$-hypergeometric function 
{\rm \cite{rf:Kaneko:96} }
\be
\qtHyperGeo 21qt{a,b}cx 
:=
\sum_{\lambda \atop \ell(\lambda)\leq M} 
P_\lambda(x) 
\prod_{(i,j)\in\lambda} {
(t^{i-1}-aq^{j-1})
(t^{i-1}-bq^{j-1})
\over
(t^{i-1}-cq^{j-1})
(1-q^{\lambda_i-j+1} t^{\lambda\dualvee_j-i})
}.
\label{eq:multiQHyper}%
\ee
}
Since $P_\lambda(x;q,t) =P_\lambda(x;q^{-1},t^{-1})$,
$\qtHyperGeo 21qt{a,b}cx $ satisfies
\be
\qtHyperGeo 21qt{a,b}cx 
=
\qtHyperGeo 21{ q^{-1} }{ t^{-1} }{ a^{-1},b^{-1} }{ c^{-1} }{ {ab\over qc}x } .
\ee
When $M=\infty$,
\be
\omega_{q,t} \qtHyperGeo 21qt{a,b}cx 
=
\qtHyperGeo 21tq{a,b}c{ {ab\over c}x },
\qquad 
M=\infty.
\ee
When $M=1$,
$\qtHyperGeo 21qt{a,b}cx $
reduces to the usual $q$-hypergeometric function
\be
\qtHyperGeo 21qt{a,b}cx
:=
\qHyperGeo 21qt{a,b}cx
:=
\sum_{n\geq 0}
x^n
\prod_{\ell=0}^{n-1}
{
(1-aq^\ell)(1-bq^\ell)
\over
(1-cq^\ell)(1-q^{\ell+1})
},
\qquad
M=1.
\label{eq:qHyperGeo}%
\ee 

In the next section we will show a relation between
our $Z_2 \left(x+{1-u\over 1-t}y\right)$ and
the 5-dimensional $SU(2)$ Nekrasov partition function.

\Section{Five-dimensional 
Nekrasov partition function} 
\label{secNek}%

Let $Q=(Q_1,\cdots,Q_\nn )$ 
and ${\massQ \pm}=({\massQ \pm_1},\cdots,{\massQ \pm_\nn} )$ 
be sets of complex parameters. 
The instanton part of the five-dimensional 
$SU(\nn)$ Nekrasov partition function
with $N_f = 2 \nn $ fundamental matters%
\footnote{
The parameters $(q,t)$ are related with those
$(\epsilon_1,\epsilon_2)$ of the $\Omega$ background  
through $(q,t)=(e^{R\epsilon_2},e^{-R\epsilon_1})$
where $R$ is the radius of the 5th dimensional circle.
The parameter $Q$ is related 
with the vacuum expectation value $a$ of the scalar fields in the  
vector multiplets
and the mass $m$ of the fundamental matter
as $Q_i=q^{a_i}$, $\pQ_i=q^{-m_i}$ and $\mQ_i=q^{-m_{\nn+i}}$.
}{ }
is written by a sum over $\nn $ Young diagrams $\lambda_i$
$(i=1,2,\cdots,\nn )$
as follows \cite{rf:Iqbal-Kashani-Poor:06}\cite{rf:Awata-Kanno:08}:%
\footnote{
In \cite{rf:Awata-Kanno:08},
there are typos in (9.4) and (9.5).
For $\alpha < \beta $, $Q_{\alpha,\beta}$ and $Q_{\alpha,\beta}'$
should be replaced with
$v^{1+{(-1)^\alpha+(-1)^\beta\over 2}}Q_{\alpha,\beta}$ and 
$v^{-1-{(-1)^\alpha+(-1)^\beta\over 2}}Q_{\alpha,\beta}'$,
respectively.
Furthermore we change the definition of $\Lambda_\alpha$ 
in \cite{rf:Awata-Kanno:08}
as follows
\be
\Lambda_\alpha
:=
\Lambda^{2\nn}
\prod_{\beta=1}^{\alpha-1}
{ Q_\beta \over \sqrt{\pQ_\beta \mQ_\beta} }
\prod_{\beta=\alpha}^\nn 
{ \sqrt{\pQ_\beta \mQ_\beta} \over Q_\beta }
.
\nonumber
\ee
}{ }
\be
Z^{\rm inst}(Q)
:=
\sum_{\{\lambda_i\}}
\prod_{i,j}
{
\sqrt{
\prod_{\epsilon=\pm}
N_{\lambda_i\bullet}(vQ_i/\massQ\epsilon_j)
N_{\bullet\lambda_i}(v\massQ\epsilon_j/Q_i)
}
\over
N_{\lambda_i\lambda_j}(Q_i/Q_j)
}
\cdot\prod_i \left(
\Lambda^2\over v
\right)^{\nn |\lambda_i| }
~~~
\label{eq:NekSymm}%
\ee
with $v:=(q/t)^\ha$ and 
\ba
N_{\lambda\mu}(Q)
:=
N_{\lambda\mu}(Q;q,t)
&:=&
\prod_{(i,j)\in\lambda} 
\left( 1 - Q\, q^{\lambda_i-j} t^{\mu\dualvee_j-i+1} \right)
\prod_{(i,j)\in\mu } 
\left( 1 - Q\, q^{-\mu_i+j-1} t^{-\lambda\dualvee_j+i  } \right)
\cr
&=&
\prod_{(i,j)\in\mu }
\left( 1 - Q\, q^{\lambda_i-j} t^{\mu\dualvee_j-i+1} \right)
\prod_{(i,j)\in\lambda}
\left( 1 - Q\, q^{-\mu_i+j-1} t^{-\lambda\dualvee_j+i  } \right).
~~~
\label{eq:NekFactor}%
\ea
Here $\lambda = (\lambda_1,\lambda_2,\cdots)$ is a Young diagram
such that $\lambda_{i} \geq \lambda_{i+1}$. 
$\lambda\dualvee $ is its conjugate Young diagram
and $|\lambda| = \sum_i \lambda_i$.
$Z^{\rm inst}(Q;\pQ;\mQ)$
is symmetric in masses ${\massQ \pm}_j$'s.
Note that $N_{\lambda\mu}(Q;q,t)$ satisfies
\be
N_{\lambda\mu}(vQ;q,t)
=
N_{\mu\lambda}(Q/v;q^{-1},t^{-1})
=
N_{\mu\dualvee\lambda\dualvee}(Q/v;t,q)
\ee
and 
\be
N_{\lambda\bullet}(vQ)
N_{\bullet\lambda}(vQ')
=
N_{\bullet\lambda}(v/Q)
N_{\lambda\bullet}(v/Q')
(QQ')^{|\lambda|}. 
\label{eq:NekQinvQ}%
\ee
Using \eqref{eq:NekQinvQ}, \eqref{eq:NekSymm} is rewritten to
the following two ways 
(double-sign corresponds):
\be
Z^{\rm inst}(Q)
=
\sum_{\{\lambda_i\}}
\prod_{i,j}
{
N_{\lambda_i\bullet}(vQ_i/\massQ\pm_j)
N_{\bullet\lambda_i}(v\massQ\mp_j/Q_i)
\over
N_{\lambda_i\lambda_j}(Q_i/Q_j)
}
\cdot\prod_i \left(
{ \Lambda^\pm_\alpha \over v^\nn } 
\right)^{|\lambda_i|}
~~~
\label{eq:NekNon}%
\ee
with
\be
{\massLambda\pm\alpha} 
:=
\Lambda^{2\nn} \prod_{j=1}^{\nn} 
\left({ \massQ\pm_j\over\massQ\mp_j }\right)^\ha.
\ee


Note that
for $\lambda$ or $\mu = (0)$, 
\be
N_{\lambda\bullet}(Q)
=
\prod_{(i,j)\in\lambda}
( 1 - Q\, q^{j-1} t^{1-i} ),
\quad
N_{\bullet\lambda}(Q)
=
\prod_{(i,j)\in\lambda}
( 1 - Q\, q^{-j} t^{i} )
\ee
and
$N_{\bullet \bullet}(Q) = 1$.
Here $\bullet$ denotes $(0)$.
%
%
Hence for a special value of $Q$,
$N_{\lambda\bullet}(1) 
= \delta_{\lambda,\bullet}$ 
and 
\be
N_{\lambda\bullet}(t) 
= 
\sum_{n\geq 0} \delta_{\lambda,n}
\prod_{\ell=0}^{n-1} 
( 1 - q^{\ell}t ),
\qquad
N_{\lambda\bullet}(1/q) 
= 
\sum_{n\geq 0} \delta_{\lambda,1^n}
\prod_{\ell=0}^{n-1} 
( 1 - t^{-\ell}/q ).
\ee
Therefore one can adjust the parameter $Q$ so that
a factor of numerator of \eqref{eq:NekSymm},
$N_{\lambda\bullet}(Q)$, 
vanishes except for $\lambda = (0)$, $(n)$ or $(1^n)$. 
Namely for some $j$,
if $vQ_i/{\massQ \pm_j} = 1$, $t$ or $q^{-1}$
then the right hand side 
of \eqref{eq:NekSymm}
is summed over only $\lambda_i = (0)$, $(n)$ or $(1^n)$ 
with $n\in\bZ_{\geq 0}$, respectively. 
%
%
Note also that
for $\lambda$, $\mu = (0)$, $(n)$ or $(1^n)$,
\ba
N_{n n}(Q)
=
\prod_{\ell=0}^{n-1}
( 1 - Q\, q^{\ell} t )
( 1 - Q\, q^{-\ell-1} ),
\quad
N_{1^n 1^n}(Q)
&=&
\prod_{\ell=0}^{n-1}
( 1 - Q\, t^{\ell+1} )
( 1 - Q\, t^{-\ell}/q ),
\cr
N_{n \bullet}(Q)
=
\prod_{\ell=0}^{n-1}
( 1 - Q\, q^{\ell} ),
\qquad
\hskip10pt
N_{\bullet n}(Q)
&=&
\prod_{\ell=0}^{n-1}
( 1 - Q\, q^{-\ell-1} t),
\cr
N_{1^n \bullet}(Q)
=
\prod_{\ell=0}^{n-1}
( 1 - Q\, t^{-\ell} ),
\qquad
N_{\bullet 1^n}(Q)
&=&
\prod_{\ell=0}^{n-1}
( 1 - Q\, t^{\ell+1}/q )
\ea
and
\be
{
N_{n\bullet}(t)
\over
N_{n n}(1)
} 
= 
\prod_{\ell=0}^{n-1} 
{1\over  1 - q^{-\ell-1}}
,
\qquad
{
N_{1^n\bullet}(1/q)
\over
N_{1^n 1^n}(1)
} 
= 
\prod_{\ell=0}^{n-1} 
{1\over  1 - t^{\ell+1}}
.
\ee


Hence one can adjust $\nn $ out of $N_f = 2\nn $ parameters ${\massQ \pm}_i$'s 
so that \eqref{eq:NekNon} reduces to 
all $\lambda_i = (0)$ but a 
$\lambda_j = (n)$ or $(1^n)$ with $n\in\bZ_{\geq 0}$ 
same as \cite{rf:Mironov-Morozov:Power}.
For example, if
$(Q_1,\cdots,Q_{\nn-1},Q_\nn) = 
v^{-1}\times(\massQ\pm_1,\cdots,\massQ\pm_{\nn-1},t\massQ\pm_\nn)$
then the right hand side 
of \eqref{eq:NekNon}
is summed over only 
$(\lambda_1,\cdots,\lambda_{\nn-1},\lambda_\nn) 
= ((0),\cdots,(0),(n))$ 
with $n\in\bZ_{\geq0}$
and thus
\ba
&&\hskip-6pt
Z^{\rm inst}
(\massQ\pm_1/v,\cdots,\massQ\pm_{\nn -1}/v,t\massQ\pm_\nn/v)
\cr
&=&
\sum_{n\geq 0}
\left(
{ {\massLambda\pm\nn} \over v^\nn }
\right)^n
{
N_{n \bullet}(vQ_\nn /\massQ\pm_\nn )
N_{\bullet n}(v\massQ\mp_\nn /Q_\nn)
\over 
N_{n n}(Q_\nn /Q_\nn )
}
\prod_{j=1}^{\nn -1}
{
N_{n \bullet}(vQ_\nn /\massQ\pm_j )
\over 
N_{n \bullet}(Q_\nn /Q_j )
}
{
N_{\bullet n}(v\massQ\mp_j /Q_\nn)
\over 
N_{\bullet n}(Q_j /Q_\nn)
}
\cr
&=&
\sum_{n\geq 0}
\left(
{  {\massLambda\pm\nn}  \over v^\nn }
\right)^n
\prod_{\ell = 0}^{n-1}
{
1-q^{-\ell} Q_\nn/v\massQ\mp_\nn
\over 
1-q^{-\ell - 1} 
}
\prod_{j=1}^{\nn -1}
{
1-q^{-\ell} Q_\nn/v\massQ\mp_j 
\over 
1- tq^{-\ell-1} Q_\nn/Q_j
}.
\ea
%
On the other hand,  if
$(Q_1,\cdots,Q_{\nn-1},Q_\nn) = 
v^{-1}\times(\massQ\pm_1,\cdots,\massQ\pm_{\nn-1},\massQ\pm_\nn /q)$
then only 
$(\lambda_1,\cdots,\lambda_{\nn-1},\lambda_\nn) 
= ((0),\cdots,(0),(1^n))$ 
contributes.
Therefore we obtain
\\
{\bf Proposition.}{\it
\ba
Z^{\rm inst}
(\massQ\pm_1/v,\cdots,\massQ\pm_{\nn -1}/v,t\massQ\pm_\nn/v)
&=&
\qHyperGeo{\nn}{\nn-1}{q^{-1}}{t^{-1}}
{
{\massQ\mp_1\over vQ_\nn },\cdots,
{\massQ\mp_{\nn}\over vQ_\nn }
}
{
{tQ_1\over qQ_\nn },\cdots,
{tQ_{\nn-1}\over qQ_\nn }
}
{
{ {\massLambda\pm\nn} \over v^\nn }
}
\cr
&=&
\qHyperGeo{\nn}{\nn-1}qt
{
v{Q_\nn \over \massQ\mp_1 },\cdots,
v{Q_\nn \over \massQ\mp_\nn }
}
{
{qQ_\nn \over tQ_1 },\cdots,
{qQ_\nn \over tQ_{\nn-1} }
}
{
v^\nn {\massLambda\mp\nn} 
}
,
\label{eq:qHyperNek}%
\cr
Z^{\rm inst}
(\massQ\pm_1/v,\cdots,\massQ\pm_{\nn -1}/v,\massQ\pm_\nn/qv)
&=&
\qHyperGeo{\nn}{\nn-1}tq
{
{\massQ\mp_1\over vQ_\nn },\cdots,
{\massQ\mp_{\nn}\over vQ_\nn }
}
{
{tQ_1\over qQ_\nn },\cdots,
{tQ_{\nn-1}\over qQ_\nn }
}
{
{ {\massLambda\pm\nn} \over v^\nn }
}
\ea
with
\be
\qHyperGeo rsqt
{a_1,\cdots,a_r}
{b_1,\cdots,b_s}
x
:=
\sum_{n\geq 0}
x^n
\prod_{\ell = 0}^{n-1}
{
(-q^\ell)^{s+1-r}
\prod_{i=1}^r (1-q^\ell a_i)
\over
(1-q^{\ell+1})
\prod_{i=1}^s (1-q^\ell b_i)
}.
\ee
}
Note that
\be
{
\qHyperGeo r{r-1}qt
{a_1,\cdots,a_r}
{b_1,\cdots,b_{r-1}}
x
}
=
{
\qHyperGeo r{r-1}{q^{-1}}t
{a_1^{-1},\cdots,a_r^{-1}}
{b_1^{-1},\cdots,b_{r-1}^{-1}}
{\widetilde x}
},
\qquad
\widetilde x
:=
{x\prod_{i=1}^r a_i \over q \prod_{i=1}^{r-1} b_i }.
\ee
%
%
When $\nn =2$,
$Z^{\rm inst}$
coincides with the $M=1$ case of the partition function $Z_2$ of the 
$q$-deformed $\beta$-ensemble 
\eqref{eq:qGMMVir}
similar to \cite{rf:Schiappa-Wyllard}
\ba
Z^{\rm inst}
(\massQ\pm_1/v,t\massQ\pm_2/v)
&=&
\qHyperGeo 21 qt
{
v{Q_2\over \massQ\mp_1 },
v{Q_2\over \massQ\mp_2 }
}
{
{qQ_2\over tQ_1 }
}
{
v^2{\massLambda\mp 2}
}
=
{ Z_2\left(x+{1-u\over 1-t}y\right) \over Z_2\left({1-u\over 1-t}y\right) }
,
\label{eq:qHyperNek2q}%
\\
Z^{\rm inst}
(\massQ\pm_1/v,\massQ\pm_2/qv)
&=&
\qHyperGeo 21 tq
{
{\massQ\mp_1\over vQ_2 },{\massQ\mp_2\over vQ_2 }
}
{
{tQ_1\over qQ_2 }
}
{
{{\massLambda\pm 2} \over v^2}
}
=
\omega_{q,t}
{ 
Z_2\left(\sum_i x_i+{1-u\over 1-t}y\right)
\over 
Z_2\left({1-u\over 1-t}y\right)
}
\label{eq:qHyperNek2t}%
\ea
with
\be
q^{s} = { \massQ\mp_1 \over vQ_2},
\quad
t^{-r} = { \massQ\mp_2 \over v Q_2},
\quad
u = {qQ_1 Q_2\over t{\massQ\mp_1} {\massQ\mp_2} },
\quad
{qx\over y} = 
{ {\massQ\mp_1} {\massQ\mp_2} \over Q_1 Q_2 }
{\massLambda\mp 2}
\ee
for \eqref{eq:qHyperNek2q} and 
\be
q^{-s} = v{Q_2 \over \massQ\mp_1},
\quad
t^{r} = v{Q_2\over \massQ\mp_2 },
\quad
u = { t{\massQ\mp_1} {\massQ\mp_2} \over qQ_1 Q_2},
\quad
{tx \over y} = 
{ Q_1 Q_2 \over {\massQ\mp_1} {\massQ\mp_2} }
{\massLambda\pm 2}
\label{eq:qGMMqHyper2t}%
\ee
for \eqref{eq:qHyperNek2t}.
%
%
In the $SU(\nn)$ case,
the Nekrasov partition function 
\eqref{eq:qHyperNek}
may coincide with 
our partition function $Z_\nn$
by using the formulas
\eqref{eq:WarnaarFormula}
and 
\eqref{eq:MacIntegral}.


\section*{Acknowledgments}


We would like thank  
H. Fuji, J. Kaneko, H. Kanno, M. Manabe, H. Ochiai, S. Odake, J. Shiraishi,
S. Warnaar and S. Yanagida 
for discussions.
The work of Y.Y. is supported in part by Grant-in-Aid for Scientific  
Research
[\#21340036] from the Japan Ministry of Education, Culture, Sports,  
Science and Technology.
The work of H.A. is supported in part by Daiko Foundation.


\Appendix{\appMac}{Macdonald polynomial}


Here we recapitulate basic properties of 
the Macdonald polynomial \cite{rf:Macdonald}.
Let $\lambda :=
(\lambda_1,\lambda_2,\cdots,\lambda_\mm)$ 
with
$\lambda_i\geq\lambda_{i+1}\geq0$ 
be a Young diagram.
$\lambda\dualvee $ is its conjugate. 
For any $\lambda$ with $\lambda_1\leq s$,
$\widehat\lambda$ is complements of $\lambda$ with respect to $(s^\mm)$,
i.e., $\widehat\lambda_i = s-\lambda_{\mm-i+1}$.
$|\lambda| := \sum_i \lambda_i$.
Let 
$x:=(x_1,\cdots,x_\mm)$
and 
$p:=(p_1,p_2,\cdots)$
with the power sum
$p_n:=p_n(x):=\sum_{i=1}^{\mm} x_i^n$.
For any symmetric function $f$ in $x$ with $\mm=\infty$, 
$f({\xp p})$ stands for the function $f$ expressed in the power sum $p$.

%
The Macdonald polynomials $P_\lambda(x):=P_\lambda(x;q,t)$ 
are degree $|\lambda|$ homogeneous symmetric polynomials in $x$ 
defined as eigenfunctions of the Macdonald operator $H$ as follows:
\ba
&&\qquad
H P_\lambda(x) =\varepsilon_\lambda P_\lambda(x),\cr
&&
H := \sum_{i=1}^\mm \prod_{j(\neq i)} {t x_i - x_j \over x_i - x_j} \cdot q^{D_{x_i}},
\qquad
\varepsilon_\lambda := \sum_{i=1}^\mm q^{\lambda_i} t^{\mm-i}
\label{eq:MacDef}%
\ea
with a normalization condition
$P_\lambda(x) = 
x_1^{\lambda_1} 
x_2^{\lambda_2} 
\cdots
x_\mm^{\lambda_\mm}
+\cdots$.
Where 
$q^{D_x}$ with  $D_x := x{\partial\over\partial x}$ 
is the $q$-shift operator such that
$q^{D_x}f(x) = f(qx)$.
Note that $P_\bullet(x) := P_{(0)}(x) = 1$.

%
Two kinds of inner products are known
in which the Macdonald polynomials are orthogonal each other.
For any symmetric functions $f$ and $g$ in $x$,
let us define inner product $\langle*,*\rangle $ 
and another one $\langle*,*\rangle'_\mm$ as follows:
\ba
\IP fg{q,t}
&:=& 
\oint\prod_{n>0}   {dp_n\over \twoPiI p_n}
\cdot
{ f({\xp{p^*}}) }\,g({\xp p}),
\qquad
p_n^* := n {1-q^n \over 1-t^n} {\partial \over \partial p_n},
\label{eq:InnerProduct}%
\\
\MacIP fg\mm
&:=& 
{1\over \mm!}
\oint\prod_{j=1}^\mm {dx_j\over \twoPiI x_j}
\cdot
\MacDelta(x)\,{f(\overline x)}\,g(x),
\qquad
\overline{x_j} := {1\over x_j}
\label{eq:AnotherInnerProduct}%
\ea
with
\be
\MacDelta(x) 
:=
\prod_{i\neq j}^\mm
\Exp{ -\sum_{n>0}{1\over n}{1-t^n\over 1-q^n}{x_j^n\over x_i^n}}
=
\prod_{i\neq j}^\mm
\prod_{\ell\geq 0} { 1-q^\ell x_j/ x_i \over 1-tq^\ell x_j/ x_i },
\qquad 
|q|<1.
\ee
Here we must treat the power sums $p_n$ as formally independent variables, 
{\it i.e.},
${\partial\over \partial p_n}\, p_m = \delta_{n,m}$ for all $n,m>0$.
%
%
The inner products of Macdonald polynomials are given by 
\ba
\IP{ P_\lambda }{ P_\mu }{q,t}
&=&
\delta_{\lambda,\mu}
\PIP{\lambda},
\qquad
\PIP{\lambda}
:=
\prod_{(i,j)\in\lambda}
{
1-q^{\lambda_i-j+1} t^{\lambda\dualvee_j-i}
\over 
1-q^{\lambda_i-j  } t^{\lambda\dualvee_j-i+1}
},
\label{eq:InnerProductValue}%
\\
\MacIP{ P_\lambda }{ P_\mu }\mm
&=& 
\delta_{\lambda,\mu}
\MacPIP{\lambda}\mm
,
\quad
{
\MacPIP{\lambda}\mm
\over 
\PIP{\lambda}
}
:=
\prod_{(i,j)\in\lambda}
{1-q^{j-1}t^{\mm-i+1} \over 1-q^{j}t^{\mm-i}}
\prod_{k=1}^\mm
{\Gamma_q(k\beta) \over \Gamma_q(\beta)\Gamma_q((k-1)\beta+1)}.
\label{eq:AnotherInnerProductValue}%
\ea
Here $\Gamma_q(x)$ is the $q$-deformed $\Gamma$ function
$
\Gamma_q(x)
:=
(1-q)^{1-x}
\prod_{\ell\geq 0} {1-q^{\ell+1}\over 1-q^{\ell+x}}
$.
%
%
%
Since the Macdonald operator is self-adjoint 
for the another inner product $\langle *,*\rangle'_\mm$, that is to say
$\langle H\,f,g\rangle'_\mm = \langle f,H\,g\rangle'_\mm$ 
(eq.\ (VI.9.4) in \cite{rf:Macdonald}),
the Macdonald polynomials are orthogonal for this product
$\langle P_\lambda,C\,P_\mu\rangle'_\mm \propto \delta_{\lambda,\mu}$ 
with an arbitrary pseudo-constant $C(x)$, i.e.,
$q^{D_{x_i}} C(x)=C(x)$.
Since 
$\qWDelta(x)/\MacDelta(x)$
is a pseudo-constant,
the other inner product replacing 
$\MacDelta(x)$ with $\qWDelta(x)$ as
\be
\qWIP fg\mm
:=
{1\over \mm!}
\oint\prod_{j=1}^\mm {dx_j\over \twoPiI x_j}
\cdot
\qWDelta(x)\,{f(\overline x)}\,g(x),
\qquad
\overline{x_j} := {1\over x_j}
\label{eq:qWInnerProduct}%
\ee
also has orthogonality, i.e., 
$
\qWIP{ P_\lambda }{ P_\mu }\mm
=
\delta_{\lambda,\mu}
\qWPIP{\lambda}\mm 
$.
%
%
Let us denote by $f({\xp{ {1-u \over 1-t} }})$
the function $f({\xp p})$
in the specialization 
$p_n := (1-u^n)/(1-t^n)$ with $u\in\bC$,
then \cite{rf:Macdonald}
\be 
P_\lambda\left({\xp{ {1-u \over 1-t } }}\right) =
\prod_{(i,j)\in\lambda}
{t^{i-1}-uq^{j-1} \over 1-q^{\lambda_i-j} t^{\lambda\dualvee_j-i+1} }.
\label{eq:uSpecialization}%
\ee

The following Cauchy formula is especially important:
\ba
\sum_\lambda
{ 1\over \PIP{ \lambda } }
P_\lambda(x;q,t) P_\lambda(y;q,t)
=
\Pi(x,y)
&:=&
\exp\left\{
\sum_{n>0}{ 1\over n}{1-t^n \over 1-q^n} p_n(x) p_n(y)
\right\}.
\label{eq:AppCauchy}%
\ea
%
With the involution $\omega_{q,t}$, 
\be
{1\over \PIP{ \lambda } }
\omega_{q,t} P_{\lambda} (x; q,t)
=
P_{\lambda\dualvee} (x; t,q), 
\label{eq:InvolMac}%
\qquad 
\omega_{q,t} (p_n) 
=
(-1)^{n-1}{1-q^n \over 1-t^n} p_n.
\ee
If we act $\omega_{q,t}$ on $x$ of $\Pi(x,y)$, 
it becomes
\be
\Pi_0(x,y)
:=
\exp\left\{
\sum_{n>0}{ (-1)^{n-1}\over n} p_n(x) p_n(y)
\right\}.
\ee



Let us denote 
a symmetric function $f$ in the set of variables
$(x_1,x_2,\cdots,y_1,y_2,\cdots)$
by
$
f\left(x,y\right)
$
or 
$
f\left(\{x,y\}\right) 
$.
Let $f_{\lambda,\mu}^\nu$ be the following fusion coefficient
\be
P_\lambda(x) P_\mu(x) =: \sum_\nu f_{\lambda,\mu}^\nu P_\nu(x),
\label{eq:FusionCoeff}%
\ee
i.e.,
$f_{\lambda,\mu}^\nu :=
{ \IP{P_\lambda P_\mu}{P_\nu}{q,t} }/{ \IP{P_\nu}{P_\nu}{q,t} }$.
Then we have
\\
{\bf Lemma.}
\be
{ P_\nu  (x,y) \over \PIP{ \nu } }
=
\sum_{\lambda,\mu\atop \lambda,\mu\subset\nu} 
{ P_\lambda(x) \over \PIP{ \lambda } }
f_{\lambda,\mu}^\nu 
{ P_\mu    (y) \over \PIP{ \mu } }. 
\label{eq:FusionFormula}%
\ee

\proof
By  the Cauchy formula \eqref{eq:AppCauchy},
\ba
\sum_\nu
{ P_\nu  (x,y) P_\nu  (z) \over \PIP{ \nu } }
&=&
\Pi(\{x,y\},z)
=
\Pi(x,z)\Pi(y,z)
\cr
&=&
\sum_{\lambda,\mu}
{ P_\lambda(x) P_\lambda(z) \over \PIP{ \lambda } }
{ P_\mu  (z) P_\mu  (y) \over \PIP{ \mu } }
\cr
&=&
\sum_{\lambda,\mu,\nu}
{ P_\lambda(x) \over \PIP{ \lambda } }
f_{\lambda,\mu}^\nu 
P_\nu(z)
{ P_\mu  (y) \over \PIP{ \mu } }
.
\ea
\qed

The fusion coefficient satisfies \cite{rf:Warnaar}
\be
f_{\lambda,\mu}^\nu
=
f_{\mu,\widehat\nu}^{\widehat\lambda}
{
\MacPIP{ \lambda }{\mm}
\over 
\MacPIP{ \nu }{\mm}
}
\ee
where $\widehat\lambda$ is the complements of $\lambda$ 
with respect to $(s^\mm)$
with $\lambda_1\leq s$.
Thus when $\nu=(s^\mm)$, 
\be
f_{\lambda,\widehat\lambda}^{(s^\mm)}
=
{
\MacPIP{ \lambda }{\mm}
\over 
\MacPIPbk{ s^\mm }{\mm}
}.
\label{eq:FusionComplement}%
\ee
Note that
$
\MacPIP{ \lambda }{\mm}
=
\MacPIP{ \widehat\lambda }{\mm}
$.
For abbreviation, let
$\PIPbk{ s^r }:= \PIP{ (s^r) }$
and 
$\PIPbk{ \widehat{n} }:= \PIP{ \widehat{(n)} }$,
then we have 
\be
{  
\PIPbk{ s^r } 
\over
\PIPbk{ n }
\PIPbk{ \widehat{n} }
}
f_{ (n),\widehat{(n)} }^{(s^r)} 
= 
\prod_{\ell=0}^{n-1}
{( 1-q^{s-\ell} )( 1-q^\ell t )
\over
( 1-q^{\ell+1} )(1-q^{s-\ell-1}t )
}.
\label{eq:FusionRow}%
\ee

For the $\mm$ variables $x:=(x_1,\cdots,x_\mm)$, 
(eq.\ (VI.4.17) in \cite{rf:Macdonald}),
\be
P_{\lambda+(s^\mm)}(x) 
=
P_\lambda(x) 
P_{(s^\mm)}(x),
\qquad 
P_{(s^\mm)}(x)
=
\prod_{j=1}^\mm x_i^s .
\label{eq:GalileanBoost}%
\ee


Let us denote the Young diagram 
decomposing into rectangles as
$\lambda = \sum_{i=1}^{\nn -1} (s_i^{r_i})$, $r_i\geq r_{i+1}$,
i.e., 
$\lambda\dualvee = (r_1^{s_1} r_2^{s_2} \cdots r_{\nn -1}^{s_{\nn -1}})$,
%
%
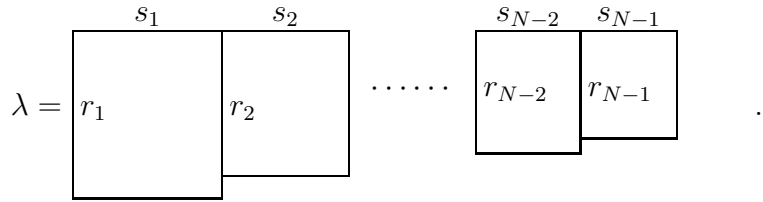
\generalYoung
%
%
\noindent
Then we have the following integral representation of the Macdonald polynomial  
\cite{rf:Awata-Odake-Shiraishi}
\be
  P_{\lambda}(x)
  =
  \prod_{a=1}^{\nn-1}
{
\PIP{ \lambda^{(a)} }
\over
r_a!
\MacPIP{ \lambda^{(a)} }{r_a}
}
\cdot 
 \oint\prod_{a=1}^{\nn-1}\prod_{j=1}^{r_a}
  {dz^a_j \over \twoPiI z^a_j}(z^a_j)^{-s_a}\cdot
  \Pi(x,z^{1})
  \prod_{a=1}^{\nn-1} \Pi(\overline{z^a},z^{a+1})\MacDelta(z^a)
\ee
with $z^\nn_i:=0$
and
$\lambda^{(1)} := \lambda$,
$\lambda^{(a)} := \sum_{i=a}^{\nn -1} (s_i^{r_i})$, 
i.e., 
${\lambda^{(a)}}\dualvee = (r_a^{s_a} r_{a+1}^{s_{a+1}} \cdots r_{\nn -1}^{s_{\nn -1}})$.
By replacing
$\MacDelta(x)$ and $z^a$ with $\qWDelta(x)$ and $p^az^a$, respectively,
we also have \cite{rf:AKOS}
\ba
P_\lambda\left(x\right)
&=&
{\qWC\lambda +}
\oint\prod_{a=1}^{\nn -1} \prod_{j=1}^{r_a} {dz^a_j\over \twoPiI z^a_j}(z^a_j)^{-s_a}
\cdot
\Pi\left(x,pz^{1}\right)
\prod_{a=1}^{\nn -1} \Pi\left(\overline{z^a},pz^{a+1}\right) \qWDelta(z^a)
\cr
&=&
{\qWC\lambda +}
\langle \alpha_{r,s}^+|\Exp{-\sum_{n>0}{{\h 1n}\over 1-q^n}\sum_{i=1}^Mx_i^n}
|\chi_{r,s}^+\rangle,
\quad
{\qWC\lambda +}
:=
  \prod_{a=1}^{\nn-1}
{
p^{-a r_a s_a}
\PIP{ \lambda^{(a)} }
\over
r_a!
\qWPIP{ \lambda^{(a)} }{r_a}
}
~~~~~
\label{eq:MacIntegral}%
\ea
with  a singular vector $|\chi_{r,s}^+\rangle$ in \eqref{eq:singular}.%
\footnote{
\eqref{eq:MacIntegral} 
can be written also by the Jackson integral 
\cite{rf:Shiraishi:Private}. 
}{ }
Acting $\omega_-\omega_+\omega_{q,t}$ on \eqref{eq:MacIntegral} gives
\be
P_{\lambda\dualvee}\left(x\right)
=
{\qWC\lambda -}
\langle \alpha_{r,s}^-|\Exp{-\sum_{n>0}{{\h 1n}\over 1-q^n}\sum_{i=1}^M(-qx_i)^n}
|\chi_{r,s}^-\rangle,
\qquad
{\qWC\lambda -}
:=
\omega_-\omega_+{ {\qWC\lambda +}\over\PIP{ \lambda } }
.
\ee


\Appendix{\appProof}{Proof of \eqref{eq:ParaSpecialComplement} }


Here we prove \eqref{eq:ParaSpecialComplement}.
We have the following formulas for the Young diagrams,
which translate the summation in squares into that in lows
\cite{rf:Awata-Kanno:08}:
\\
{\bf Lemma.}
\be
(1-q)\sum_{(i,j)\in\lambda} q^{j-1} t^{1-i} 
=
\sum_{i=1}^{r} \left(1-q^{\lambda_i}\right) t^{1-i},
\qquad
r\geq\ell(\lambda),
\label{eq:partitionFormulaI}
\ee
\be
(1-q)\sum_{(i,j)\in\mu } q^{\lambda_i-j} t^{\mu\dualvee_j-i} 
=
\left[
\sum_{i=1}^{r}
\sum_{j=i}^{r}
-t^{-1}
\sum_{i=1}^{r}
\sum_{j=i+1}^{r + 1}
\right]
q^{\lambda_i-\mu_j} t^{j-i},
\qquad
r\geq\ell(\mu).
\label{eq:partitionFormulaII}
\ee

In the following let us denote 
by
$\eqref{eq:partitionFormulaI}(\lambda;q,t)$ and
$\eqref{eq:partitionFormulaII}(\lambda,\mu)$
the equations
\eqref{eq:partitionFormulaI} and 
\eqref{eq:partitionFormulaII},
respectively. 
Using these we obtain
\\
{\bf Lemma.}~{\it
For any integer $r\geq\ell(\lambda),\ell(\mu)$,
\ba
\sum_{i=1}^{r} 
{
q^{\mu_i} t^{r-i}-q^{s-\lambda_i}t^{i-1}
\over
1-q
}
&=&
\sum_{(i,j)\in(s^r)} 
q^{j-1} t^{i-1} 
-
\sum_{(i,j)\in\lambda} 
q^{s-j} t^{i-1} 
-
\sum_{(i,j)\in\mu} 
q^{j-1} t^{r-i} 
\label{eq:partitionFormulaIb}
\\
&=&
\sum_{(i,j)\in\widehat\mu } 
q^{\widehat\lambda_i-j} t^{\widehat\mu\dualvee_j-i} 
-
\sum_{(i,j)\in\lambda } 
q^{\mu_i-j} t^{\lambda\dualvee_j-i} 
.
\label{eq:partitionFormulaIIb}
\ea
}
\proof
First,
$
q^s\times\eqref{eq:partitionFormulaI}(\lambda;q^{-1},t^{-1})
-t^{r-1}\times\eqref{eq:partitionFormulaI}(\mu;q,t)
$
and 
$
\sum_{i=1}^r 
{( t^{r-1} - q^st^{i-1} )/(1-q) }
=
\sum_{i=1}^r 
\sum_{j=1}^s q^{j-1}t^{i-1} 
$
gives \eqref{eq:partitionFormulaIb}.
Next, by
$
\eqref{eq:partitionFormulaII}(\widehat\lambda,\widehat\mu)
$
and
$
\widehat\lambda_i := s - \lambda_{r-i+1}
$
with
$\lambda_0 := 0$,
\ba
(1-q)\sum_{(i,j)\in\widehat\mu } 
q^{\widehat\lambda_i-j} t^{\widehat\mu\dualvee_j-i} 
&=&
\left[
\sum_{i=1}^{r}
\sum_{j=i}^{r}
-t^{-1}
\sum_{i=1}^{r}
\sum_{j=i+1}^{r + 1}
\right]
q^{\widehat\lambda_i-\widehat\mu_j} t^{j-i}
\cr
&=&
\left[
\sum_{i=1}^{r}
\sum_{j=i}^{r}
-t^{-1}
\sum_{i=0}^{r-1}
\sum_{j=i+1}^{r}
\right]
q^{\mu_i-\lambda_j} t^{j-i}
\\
&=&
\left[
\sum_{i=1}^{r}
\sum_{j=i}^{r}
-t^{-1}
\left(
\sum_{i=1}^{r}
\sum_{j=i+1}^{r+1}
-
\sum_{i=1}^{r}
\sum_{j}
\delta_{j,r+1}
+
\sum_{j=1}^{r}
\sum_{i}
\delta_{i,0}
\right)
\right]
q^{\mu_i-\lambda_j} t^{j-i}
.
\nonumber
\ea
Thus
$
\eqref{eq:partitionFormulaII}(\widehat\lambda,\widehat\mu)
-\eqref{eq:partitionFormulaII}(\mu,\lambda)
$
gives \eqref{eq:partitionFormulaIIb}.
\qed

Note that 
${\widehat\mu}\dualvee \neq\widehat{\mu\dualvee}$.
{}From \eqref{eq:partitionFormulaIb} and
\eqref{eq:partitionFormulaIIb}
and their $\mu=\widehat\lambda$ cases
we have
\ba
\sum_{(i,j)\in(s^r)} 
q^{j-1} t^{1-i}
&=&
\sum_{(i,j)\in\lambda} 
q^{s-j} t^{i-r} 
+
\sum_{(i,j)\in\widehat\lambda } 
q^{j-1} t^{1-i} 
,
\label{eq:partitionFormulaIc}
\\
\sum_{(i,j)\in(s^r)} 
q^{j-1} t^{i-1}
&=&
\sum_{(i,j)\in\lambda} 
q^{s-j} t^{i-1} 
+
\sum_{(i,j)\in\mu} 
q^{j-1} t^{r-i}
+
\sum_{(i,j)\in\widehat\mu } 
q^{\widehat\lambda_i-j} t^{\widehat\mu\dualvee_j-i} 
-
\sum_{(i,j)\in\lambda } 
q^{\mu_i-j} t^{\lambda\dualvee_j-i} 
.~~~~~ 
\label{eq:partitionFormulaIIc}
\ea
For any equation $f(q,t)=0$ 
we define a 
mapping by
$
\Exp{-\sum_{n>0}{1\over n} f(q^n,t^n)} = 0
$.
Acting this mapping on
$u\times\eqref{eq:partitionFormulaIc}$ and 
\eqref{eq:partitionFormulaIIc}
gives
\be
\prod_{(i,j)\in(s^r)} 
(1-u q^{j-1} t^{1-i})
=
\prod_{(i,j)\in\lambda} 
(1-uq^{s-j} t^{i-r} )
\prod_{(i,j)\in\widehat\lambda } 
(1-uq^{j-1} t^{1-i}) 
,
\label{eq:partitionFormulaId}
\ee
\be
\prod_{(i,j)\in(s^r)} 
(1-q^{j} t^{i-1} )
=
\prod_{(i,j)\in\lambda} 
(1-q^{s-j+1} t^{i-1} )
\prod_{(i,j)\in\mu} 
(1-q^{j} t^{r-i})
{
\prod_{(i,j)\in\widehat\mu } 
(1-q^{\widehat\lambda_i-j+1} t^{\widehat\mu\dualvee_j-i} )
\over
\prod_{(i,j)\in\lambda } 
(1-q^{\mu_i-j+1} t^{\lambda\dualvee_j-i} )
}
. 
\label{eq:partitionFormulaIId}
\ee
Thus
\be
\prod_{(i,j)\in(s^r)} 
{
(1-u q^{j-1} t^{1-i})
\over
(1-q^{j} t^{i-1} )
}
=
\prod_{(i,j)\in\lambda} 
{
(1-uq^{s-j} t^{i-r} )
\over
(1-q^{s-j+1} t^{i-1} )
}
{
\prod_{(i,j)\in\widehat\lambda } 
(1-uq^{j-1} t^{1-i}) 
\over
\prod_{(i,j)\in\mu} 
(1-q^{j} t^{r-i})
}
{
\prod_{(i,j)\in\lambda } 
(1-q^{\mu_i-j+1} t^{\lambda\dualvee_j-i} )
\over
\prod_{(i,j)\in\widehat\mu } 
(1-q^{\widehat\lambda_i-j+1} t^{\widehat\mu\dualvee_j-i} )
}
. 
\ee
When $\mu=\lambda$,
\be
\prod_{(i,j)\in(s^r)} 
{
(1-u q^{j-1} t^{1-i})
\over
(1-q^{j} t^{i-1} )
}
=
\prod_{(i,j)\in\lambda} 
{
(1-uq^{s-j} t^{i-r} )
\over
(1-q^{s-j+1} t^{i-1} )
}
{
(1-q^{\mu_i-j+1} t^{\lambda\dualvee_j-i} )
\over
(1-q^{j} t^{r-i})
}
\prod_{(i,j)\in\widehat\lambda } 
{
(1-uq^{j-1} t^{1-i}) 
\over
(1-q^{\widehat\lambda_i-j+1} t^{\widehat\lambda\dualvee_j-i} )
}
. 
\ee
\eqref{eq:uSpecialization} and 
\eqref{eq:InnerProductValue}
completes the proof of \eqref{eq:ParaSpecialComplement}.

\newpage
\Appendix{\appInteger}{Relation with Kaneko's integral formula}


When $\beta\in\bN$
we can use Kaneko's integral formula \cite{rf:Kaneko:97}.
Let us define the following another kernel,
which has the same $q=1$ limit with $\qWDelta(z)$, 
\ba
\qHyperDelta(z) 
&:=&
\prod_{i<j}
\Exp{-\sum_{n>0}{1\over n}
{ t^{n}-t^{-n} \over q^{n\over 2}-q^{-{n\over 2}} }
q^{n\over 2}
{z_j^n\over z_i^n}}
\cdot\prod_{i=1}^r z_i^{(r+1-2i)\beta}
\cr
&=&
\prod_{i<j}
\prod_{\ell\geq 1}
{1- q^\ell z_j/t z_i \over 1-q^\ell t  z_j/z_i }
\cdot\prod_{i=1}^r z_i^{(r+1-2i)\beta},
\qquad |q|<1.
\label{eq:qHyperDelta}%
\ea
In this section, we concentrate on the case of $\beta\in\bN$ with $t=q^\beta$.
Then
\ba
\Pi(z,w) 
&=& \prod_{i,j} \prod_{\ell=0}^{\beta-1}
(1-q^\ell z_i w_j )^{-1},
\cr
\qWDelta(z) 
&=&
\prod_{i<j}
\prod_{\ell=0}^{\beta-1}
(1-q^{-\ell} z_j/z_i )(1- q^\ell  z_j/z_i )
\cdot\prod_{i=1}^r z_i^{(r+1-2i)\beta}
\cr
&=&
\left( -q^{{1-\beta\over 2}} \right)^{ {\beta r (r-1)\over 2 }}
\prod_{i<j}
\prod_{\ell=0}^{\beta-1}
(1-q^\ell z_i/z_j )(1- q^\ell  z_j/z_i )
,
\cr
\qHyperDelta(z) 
&=&
\prod_{i<j}
\prod_{\ell=1-\beta}^{\beta}
(1- q^\ell  z_j/z_i )
\cdot\prod_{i=1}^r z_i^{(r+1-2i)\beta}
\cr
&=&
\left( -q^{{1-\beta\over 2}} \right)^{ {\beta r (r-1)\over 2 }}
\prod_{i<j}
\prod_{\ell=0}^{\beta-1}
(1-q^\ell z_i/z_j )(1- q^{\ell+1}  z_j/z_i )
.
\ea


Let $z:=(z_1,\cdots,z_r)$
and $x:=(x_1,\cdots,x_M)$.
We have the following Kaneko's integral formula 
for the multivariate $q$-hypergeometric function
in \eqref{eq:multiQHyper}
\\
{\bf Lemma.}
\cite{rf:Kaneko:97}\cite{rf:Zeilberger:94}
{\it
\be
\oint \prod_{j=1}^{r} {dz_j\over \twoPiI z_j}
\prod_{\ell=0}^{a-1} (1-q^\ell z_j)
\prod_{\ell=0}^{b-1} (1-q^{\ell+1}/z_j)
\prod_{k=1}^M (1-z_j x_k)
\cdot
\qHyperDelta(z) 
=
\qtHyperGeo 21tq 
{ t^{-r},q^b }{ q^{-a-1}t^{1-r} }{ q^{-a}tx }
c_{a,b,r}^{(q,t)}
\label{eq:kanekoFormula}%
\ee
with
\ba
c_{a,b,r}^{(q,t)}
&:=&
\prod_{j=0}^{r-1}
{
\prod_{\ell=0}^{a+b+j\beta} (1-q^{\ell+1})
\prod_{\ell=0}^{(j+1)\beta} (1-q^{\ell+1})
\over
\prod_{\ell=0}^{a+j\beta} (1-q^{\ell+1})
\prod_{\ell=0}^{b+j\beta} (1-q^{\ell+1})
\prod_{\ell=0}^{\beta} (1-q^{\ell+1})
}.
\ea
}
Note that the  right hand side 
of \eqref{eq:kanekoFormula}
is summed over all Young diagrams
$\lambda$ with $\lambda_1\leq r$ and $\ell(\lambda)\leq b$.

 
For any symmetric Laurent polynomial $f(z)$, we have
\\
{\bf Lemma.}
\cite{rf:Warnaar}
\be
\oint \prod_{j=1}^{r} {dz_j\over \twoPiI z_j}\cdot
\qWDelta(z) f(z)
=
r! \prod_{\ell=1}^r {1-t \over 1-t^\ell}
\oint \prod_{j=1}^{r} {dz_j\over \twoPiI z_j}\cdot
\qHyperDelta(z) f(z).
\label{eq:qWtoqH}%
\ee
This follows from 
\be
\qHyperDelta(z)
=
\qWDelta(z) 
\prod_{i<j} {1-tz_j/z_i \over 1-z_j/z_i}
\ee
and 
\be
\sum_{\omega\in \cS_r} \omega\left(
\prod_{i<j} {1-tz_j/z_i \over 1-z_j/z_i}
\right)
=
\prod_{i=1}^r {1-t^i\over 1-t}.
\ee


Let $x:=(x_1,\cdots,x_M)$.
When 
\be
-
{ t^{n\over 2}-t^{-{n\over 2}} \over q^{n\over 2}-q^{-{n\over 2}} }
p_n 
=
\sum_{k=1}^M x_k^n
+
{ q^{c{n\over 2}}-q^{-c{n\over 2}} \over q^{n\over 2}-q^{-{n\over 2}} }  
q^{-{n\over 2}(1+2s-c)} 
,
\ee
the partition function $Z_2$ in \eqref{eq:Ztwo} is
\be
Z_2 
=
\oint \prod_{j=1}^{r} {dz_j\over \twoPiI z_j}
z_j^{-s}
\prod_{\ell\geq 0}
{1-q^{\ell-s} z_j \over 1-q^{\ell+c-s} z_j }
\prod_{k=1}^M (1-z_j x_k)
\cdot
\qWDelta(z)
=:
Z_2^{(x)}.
\label{eq:ZoneGamma}%
\ee
When $c\in\bN$,
this reduces to
\be
Z_2^{(x)}
=
(-1)^s q^{-{s(s-1)\over 2}} 
\oint \prod_{j=1}^{r} {dz_j\over \twoPiI z_j}
\prod_{\ell=0}^{c-s-1}
(1-q^\ell z_j)
\prod_{\ell=0}^{s-1}
(1-q^{\ell+1}/z_j)
\prod_{k=1}^M (1-z_j x_k)
\cdot
\qWDelta(z).
\ee
Therefore, from 
\eqref{eq:kanekoFormula} and 
\eqref{eq:qWtoqH}
we obtain
\\
{\bf Proposition.}
\be
{ Z_2^{(x)} \over Z_2^{(0)} }
=
\qtHyperGeo 21tq 
{ t^{-r},q^{s} }{ q^{s-c-1}t^{1-r} }{ q^{s-c}tx }.
\label{eq:qGMMVirInteger}%
\ee
%
%
Note that 
the right hand side of \eqref{eq:qGMMVirInteger} is nothing but 
that of \eqref{eq:qGMMdualVir} with $y=q^{-s}$ and $u=t^c$.
%
Thus when $\nn =2$ and $M=1$,
\eqref{eq:qGMMVirInteger}
coincides with the 5-dimensional $SU(2)$ 
Nekrasov partition function 
\eqref{eq:qHyperNek2t}
\be
{ Z_2^{(x)} \over Z_2^{(0)} }
=
Z^{\rm inst} 
(\massQ\pm_1/v,\massQ\pm_2/qv)
\ee
with \eqref{eq:qGMMqHyper2t}.

\newpage
\Appendix{\appSelberg}{Relation with Jackson integral}


In this section we assume $0<q<1$.
In this article 
${\oint{dz\over \twoPiI z} f(z)}$ 
denotes the constant term in $f(z)$, i.e.,
${\oint{dz\over \twoPiI z} }\sum_{n\in\bZ} f_n z^n 
:=\CT{z}{ \sum_{n\in\bZ} f_n z^n }:= f_0
$.
But to define the $q$-deformed $\beta$-ensemble \eqref{eq:qGMM},
one can replace it by the Jackson integral 
which may have more natural $q\rightarrow 1$ limit.
The Jackson integral is defined by 
\be 
\int_0^1 {d_q z\over z} f(z) := (1-q) \sum_{n\geq 0} f(q^n),
\qquad
\int_0^\infty {d_q z\over z} f(z) := (1-q) \sum_{n\in\bZ} f(q^n).
\ee
For a Laurent polynomial $f$, the relation between 
the Jackson integral and the constant term map is \cite{rf:Askey}
\be
\CT z {f(z)} = 
\lim_{\epsilon\rightarrow 0}
{1-q^\epsilon \over 1-q}
\int_0^1 {d_q z\over z} z^\epsilon f(z).
\ee

For some special cases,
one can calculate the partition function of the 
$q$-deformed $\beta$-ensemble \eqref{eq:qGMM}
by using following formulas:
For $\beta\in\bZ_{\geq 0}$, $\Re(x)>0$ and $y\neq 0,-1,-2,\cdots$
\cite{rf:Kaneko:96},
\ba
&&\hskip-12pt
\int_0^1 
\prod_{j=1}^{r} {d_q z_j\over z_j} z_j^{x}
\prod_{\ell\geq0} {1-q^{\ell+1}z_j \over 1-q^{\ell+y}z_j }
\cdot
\prod_{i<j}
\prod_{\ell=1}^{2\beta} ( z_i-q^{\ell}z_j/t ) 
\cdot
\prod_{i,j} ( 1-z_i x_j )
\cr
&=&
t^{A_r}
\prod_i 
{ \Gamma_q(i\beta +1) \over \Gamma_q(\beta+1) }
{ \Gamma_q( x+(r-i)\beta ) \Gamma_q( y+(r-i)\beta )\over \Gamma_q( x+y+(2r-i-1)\beta ) }
\qtHyperGeo 21 tq { t^{-r}, q^{-x} t^{1-r} } { q^{-x-y} t^{2-2r} } {q^{-y} t x }
~~~
\ea
with
\be
A_r := { r(r-1)\over 2 }x + { r(r-1)(r-2)\over 3 }\beta. 
\ee
And also for $\beta\in\bN$ and $\Re(x)>-\lambda_r$
(Cor. 1.6 in \cite{rf:Warnaar}),%
\footnote{
The first factor 
$\prod_{\ell=0}^\beta  ( 1-q^{\ell+1}z_j  )$
can be generalized to 
$\prod_{\ell\geq0} 
{(1-q^{\ell+1}z_j )/( 1-q^{\ell+y}z_j )}$
\cite{rf:Warnaar:Private}. 
}{ }
\ba
&&\hskip-12pt
\int_0^1 
\prod_{j=1}^{r} {d_q z_j\over z_j} z_j^{x} 
\prod_{\ell=0}^\beta  ( 1-q^{\ell+1}z_j  )
\cdot
\prod_{i<j}
\prod_{\ell=1}^{2\beta} ( z_i-q^{\ell}z_j/t ) 
\cdot
\prod_{i,j} ( 1-z_i x_j )
\cdot
P_\lambda(z;q,t)
\cr
&=&
t^{A_r}
P_\lambda(1,t,\cdots,t^{r-1};q,t)
\prod_i 
{ \Gamma_q(i\beta +1)  \Gamma_q(i\beta) \over \Gamma_q(\beta+1) }
{ \Gamma_q( x+\lambda_i+(r-i)\beta ) \over \Gamma_q( x+\lambda_i+(2r-i)\beta ) }
\cr
&\times&
\qtHyperGeo {r+1}r tq 
{ t^{-r},
  q^{-x-\lambda_1} t^{2-2r},q^{-x-\lambda_2} t^{3-2r},\cdots,q^{-x-\lambda_r} t^{1-r}} 
{ q^{-x-\lambda_1} t^{1-2r},q^{-x-\lambda_2} t^{2-2r},\cdots,q^{-x-\lambda_r} t^{  -r} } 
{ x }.
\label{eq:WarnaarFormula}%
\ea

Note that If $\beta\notin\bZ$ then for any regular function $f(z)$ in $|z|\leq 1$,
using the pseudo constant $F(z)$ in \eqref{eq:qWDeltaSC}, 
we can rewrite the contour integral to the Jackson integral
\be
\oint_{|z|=1} {dz\overtwoPiI } 
F(z) z^{-\beta} f(z) 
=
\sum_{n\geq 0} q^{n(1-\beta)} f(q^n)
{\rm res}_{z=1} F(z) 
=
{ 1\over \Gamma_q(\beta) \Gamma_q(1-\beta)}
\int_0^1 {d_qz } z^{-\beta} f(z).
\ee




\Appendix{\appQone}{Four-dimensional case}



\subsection{$q\rightarrow 1$ limit}


Here we give an example when $q=1$, i.e., four-dimensional case
\cite{rf:AMOS:Note}.
Let us change the normalization of bosons by%
\footnote{
Then
$
{ {\h in}^{{\rm old}}
\over q^{n\over 2\ell} - q^{-{n\over 2\ell}} } 
=
{1\over n}
{
\sqrt{ { \qpint\beta{q^n} } }
\over 
{ \qpint{1/\ell}{q^n} } 
}
{ {\h in}^{{\rm new}} }
$
,
$
{ {\h in}^{{\rm old}}\over t^{n\over 2 k  } - t^{-{n\over 2 k  }} } 
=
{1\over n}
{ {\h in}^{{\rm new}}
\over 
{ \qpint{1/k}{t^n} }\sqrt{ { \qpint\beta{q^n} } }
}
$
and
$
{
(u^{{n\over 2}}-u^{-{n\over 2}})
{\h in}^{{\rm old}}
\over 
(q^{{n\over 2}}-q^{-{n\over 2}})
(t^{{n\over 2}}-t^{-{n\over 2}})
}
= 
{ {\qpint\gamma{t^n}}\over n }
\sqrt{ {\qpint\beta{q^n}} }
{ {\h in}^{{\rm new}} }
$
for $u=t^\gamma$.
}{ }
\be
{\h in}^{{\rm old}}
=
\sqrt{ {
( q^{{n\over 2}}-q^{-{n\over 2}} )( t^{{n\over 2}}-t^{-{n\over 2}} ) 
\over n^2
} }
{\h in}^{{\rm new}}
=
{ q^{{n\over 2}}-q^{-{n\over 2}} \over n }
\sqrt{ { \qpint\beta{q^n} } }
{\h in}^{{\rm new}}
,
\qquad n\neq0
\ee
with 
${\h i0}^{{\rm old}}={\h i0}^{{\rm new}}$ and
${\Qh i}^{{\rm old}}={\Qh i}^{{\rm new}}$ 
unchanged.
Using the notation defined in the next subsection,
let $q=e^{\hbar/\sqrt\beta}$,
$
{\h in}^{{\rm new}} 
= 
{\bosonIP{\vec h^i}{\vec a_n}{} }
+\cO(\hbar)$
and
$
{\Qh i}^{{\rm new}} 
= 
{\bosonIP{\vec h^i}{\vec Q}{} }
+\cO(\hbar)$.%
\footnote{
$\cO(\hbar)$ is a linear combination of 
${\bosonIP{\vec h^j}{\vec a_n}{} }$'s and
${\bosonIP{\vec h^j}{\vec Q}{} }$'s
with $j=1,\cdots,\nn$
\cite{rf:AKOS:W}.
}{ }
Then
\be
p^{D_z} - \Lambda_i (zp^{1-i})
=
-\hbar z^{{\nn +1 \over 2} -i+1}
( \alpha_0\partial_z + { \bosonIP{\vec h^i}{\partial_z\vec\phi}{(z)} } )
z^{-{\nn +1 \over 2} +i}
+\cO(\hbar^2)
\label{eq:QoneMiura}%
\ee
with
$\alpha_0 := \sqrt\beta-{1/\sqrt\beta}$ and
$\partial_z := {\partial\over \partial z}$.
By letting $\hbar\rightarrow 0$
we obtain the four-dimensional case.
Note that $\lim_{q\rightarrow 1}{\qpint nq} = n$.


\subsection{{\W} algebra}





Let $\{\vec{e}_i\}_{i=1}^\nn$ to be
an orthonormal basis, i.e., $\vec{e}_i\cdot\vec{e}_j=\delta_{ij}$.
The weight space of $A_{\nn-1}$ is the hyper-surface perpendicular to
$\sum_{i=1}^\nn\vec{e}_i$.
The weights of the vector representation $\vec{h}^i$, 
the simple roots $\vec{\alpha}^a$ and 
the fundamental weights $\vec{\Lambda}^a$ 
for  $i=1,\cdots,\nn$ and $a=1,\cdots,\nn-1$
are given by
$\vec{h}^i:=\vec{e}_i-\sfrac{1}{\nn}\sum_{j=1}^\nn\vec{e}_j$,
$\vec{\alpha}^a:=\vec{h}^a-\vec{h}^{a+1}$ and 
$\vec{\Lambda}^a:=\sum_{i=1}^a\vec{h}^i$.
Their inner-products are
\ba
{\bosonIP{\vec h^i}{\vec h^j}{ } }
&=&
\delta^{i,j}-{1\over \nn},
\qquad\hskip60pt
{\bosonIP{\vec\alpha^a}{\vec\alpha^b}{ } }
=
C^{ab}:=2\delta^{a,b}-\delta^{a-1,b}-\delta^{a+1,b}, 
\cr
{\bosonIP{\vec h^i}{\vec\alpha^b}{ } }
&=&
 B^{i,b}
:=
\ddelta_{i,b} -\ddelta_{i-1,b},
\qquad\hskip12pt
{\bosonIP{\vec\alpha^a}{\vec\Lambda^b}{ } }
=
\delta^{a,b}, 
\\
{\bosonIP{\vec h^i}{\vec\Lambda^b}{ } }
&=&
 A^{i,b}
:=
\theta(i\leq b) - {b\over \nn},
\qquad
{\bosonIP{\vec\Lambda^a}{\vec\Lambda^b}{ } }
=
(C^{-1})^{ab}=\min(a,b)\Bigl(1-{\max(a,b)\over N}\Bigr).
~~~~~
\nonumber
\ea
We define the boson field in the weight space by
\ba
  &&
 \vec{\phi}(z)
:=
\vec{Q}+\vec{a}_0\log z
  -\sum_{n\neq 0}\sfrac{1}{n}\vec{a}_nz^{-n},
\qquad
[
{\bosonIP{\vec\alpha^a}{\vec Q  }{} },
{\bosonIP{\vec\Lambda^b}{\vec Q  }{} }
]
= 0,
\cr
  &&
[
{\bosonIP{\vec\alpha^a}{\vec a_n}{} },
{\bosonIP{\vec\Lambda^b}{\vec a_m}{} }
]
=
n\delta^{a,b}\delta_{n+m,0},
\qquad
[
{\bosonIP{\vec\alpha^a}{\vec a_0}{} },
{\bosonIP{\vec\Lambda^b}{\vec Q  }{} }
]
=
\delta^{a,b}.
\ea
Then for any vectors $\vec u$ and $\vec v$ in the weight space,
\be
{\bosonIP{\vec u}{\vec\phi}{(z)} }
{\bosonIP{\vec v}{\vec\phi}{(w)} }
=
{\bosonIP{\vec u}{\vec v}{ } }\log(z-w)+
\NP
{\bosonIP{\vec u}{\vec\phi}{(z)} }
{\bosonIP{\vec v}{\vec\phi}{(w)} }
\NP.
\ee





{}From 
\eqref{eq:QoneMiura}
the {\W} generators $\Wcl^i(z)$ can be defined by
\be
\sum_{i=0}^\nn \Wcl^i(z)(\alpha_0 \partial_z)^{\nn-i}
:=
\NP 
(\alpha_0\partial_z + { \bosonIP{\vec h^1}{\partial_z\vec\phi}{(z)} } )
(\alpha_0\partial_z + { \bosonIP{\vec h^2}{\partial_z\vec\phi}{(z)} } )
\cdots
(\alpha_0\partial_z + { \bosonIP{\vec h^\nn}{\partial_z\vec\phi}{(z)} } )
\NP
.
\ee
Note that $\Wcl^i(z)\neq\lim_{\hbar\rightarrow 0} W^i(z)$.
$\Wcl^2(z)$ is the Virasoro generator
with the central charge 
$c =
  \nn-1-12\alpha_0^2\vec{\rho}^{\,2}$, 
\be
  -\Wcl^2(z)=
  \sfrac{1}{2}\NP ( \partial\vec{\phi}(z)\cdot\partial\vec{\phi}(z) )\NP
  +\alpha_0 (\vec{\rho}\cdot\partial^2\vec{\phi}(z) )
\ee
where
$\vec{\rho}$ is the half-sum of positive roots,
$\vec{\rho}:=\sum_{a=1}^{\nn-1}\vec{\Lambda}_a$, and
$\vec{\rho}^{\,2}=\frac{1}{12}\nn(\nn^2-1)$.
%
%
Screening currents \eqref{eq:SCDef}
and primary fields 
\eqref{eq:uqtPrimary},
\eqref{eq:primaryPM} and 
\eqref{eq:primaryellk}
are now
\be
S^a_\pm(z) := 
\NP e^{\pm\sqrt\beta^{\pm1}{ \bosonIP{\vec\alpha^a}{\vec\phi}{(z)} } } \NP,
\qquad
\VO^a_\pm(z) := 
\NP e^{\mp\sqrt\beta^{\pm1}{ \bosonIP{\vec\Lambda^a}{\vec\phi}{(z)} } } \NP,
\ee
\be
\VO^a_{\ell+1,k+1}(z) := 
\NP e^{(-\ell\sqrt\beta + k/\sqrt\beta ) 
{ \bosonIP{\vec\Lambda^a}{\vec\phi}{(z)} } } \NP
,\qquad
\VO^a_\gamma(z) := 
\NP e^{-{\gamma\sqrt\beta} { \bosonIP{\vec\Lambda^a}{\vec\phi}{(z)} } } \NP
.
\ee


\subsection{$\beta$-ensemble}





The vertex operator in \eqref{eq:VnDef} 
is now
\be
V_\nn
:=
\prod_{a=1}^{\nn -1} 
\Exp{\sqrt{\beta}\sum_{n>0}{1\over n}
{\bosonIP{\vec\Lambda^a}{\vec a_n}{} } 
p^{(a)}_n }.
\ee
Then, as \eqref{eq:isoDef},  
$\langle \alpha|V_\nn$ 
defines the isomorphism by
\be
\sqrt{\beta}
\ p^{(a)}_n 
\langle \alpha|V_\nn
=
\langle \alpha|V_\nn
{\bosonIP{\vec\alpha^a}{\vec a_{-n}}{} },
\qquad 
{n\over \sqrt{\beta} }
\deldel{ p^{(a)}_n } 
\langle \alpha|V_\nn
=
\langle \alpha|V_\nn
{\bosonIP{\vec\Lambda^a}{\vec a_{n}}{} }
\ee
for $n>0$ and 
$
{\vec\alpha}
\langle \alpha|V_\nn
=\langle \alpha|V_\nn
{\vec a_{0}}
$.
Here
$\langle \alpha|$
is an abbreviation of $\langle \vec\alpha|$.
%
%
As \eqref{eq:LinearMap}
the vector
$| S_{r,s}^+\rangle$
also defines another linear map by
\be
{\bosonIP{\vec\Lambda^a}{\vec a_{n}}{} }
| S_{r,s}^+\rangle
=
| S_{r,s}^+\rangle
\sqrt{\beta}
\sum_{k=1}^{r_a} (z^a_k)^n
,
\qquad
n>0.
\ee





The partition function $Z_\nn$,
the potential $W(z^a,z^{a+1})$ and 
the effective action $S_{\rm eff}$ 
in 
\eqref{eq:qGMM},
\eqref{eq:qSuperPotentioal} and 
\eqref{eq:Seff}, respectively, are now
\ba
Z_\nn
&:=& 
\oint\prod_{a=1}^{\nn -1}\prod_{j=1}^{r_a} {dz^a_j \over \twoPiI z^a_j}
(z^a_j)^{-s_a}
\Exp{\beta
\sum_{n>0}{1\over n}
(z^a_j)^n p^{(a)}_n }
\cdot
\qWDelta(z^a)
\Pi\left(\overline{z^a},z^{a+1}\right) 
,
\cr
W(z^a,z^{a+1})
&:=& 
\sum_{i=1}^{r_a}
\left(
\beta
\sum_{n>0}{1\over n}
(z^a_i)^n p^{(a)}_n 
-
\beta
\sum_{j=1}^{r_{a+1}}
\log\left(1-{ z^{a+1}_j \over z^a_i }\right) 
-
(s_a+1)
\log z^a_i
\right)
,
\\
S_{\rm eff}
&:=&
\sum_{a=1}^{\nn-1}
\left(
W(z^a,z^{a+1})
+
2\beta
\sum_{i<j}
\log\left(1-{ z^{a}_j \over z^{a}_i }\right)
+
\beta
\sum_{i=1}^{r_a}
(r_a+1-2i)
\log z^{a}_i
\right)
\nonumber
\ea
with $z^\nn := 0$ and 
\ba
\Pi(z,w) 
&:=& \prod_{i,j} 
\Exp{ \beta
\sum_{n>0}{1\over n}
z_i^n w_j^n }
= \prod_{i,j} 
( 1- z_i w_j )^{-\beta},
\\
\qWDelta(z) 
&:=&
\prod_{i<j}
\Exp{-2\beta
\sum_{n>0}{1\over n}
{z_j^n\over z_i^n}}
\cdot\prod_{i=1}^r z_i^{(r+1-2i)\beta}
=
\prod_{i<j}
(1- z_j/z_i)^{\beta}
( z_i/z_j-1)^{\beta}
.~~~
\ea
$Z_\nn$ is written as \eqref{eq:ZbyMac}
by the Jack polynomial $P_\lambda(x):=P_\lambda(x;\beta)$
defined in appendix {\appQone}.6.
%
%
The saddle point condition is
${\partial S_{\rm eff} \over \partial z^a_k } = 0 $
with
\ba
z^a_k 
{\partial S_{\rm eff} \over \partial z^a_k }
=
\beta
\sum_{n>0}
(z^a_k)^n 
p^{(a)}_n 
&+&
\beta\log
{
\prod_{i<k}
\left(1-{ z^{a}_k / z^{a}_i }\right)^2
\over
\prod_{j>k}
\left(1-{ z^{a}_j / z^{a}_k }\right)^2
}
{
\prod_{j=1}^{r_{a+1}}
\left(1-{  z^{a+1}_j / z^{a}_k }\right)
\over
\prod_{i=1}^{r_{a-1}}
\left(1-{ z^{a}_k / z^{a-1}_i }\right)
}
\cr
&+&
((r_a+1-2k)\beta - s_a-1)
.
\ea




As \eqref{eq:ConstraintMiura}
let us define ${\cphi}$ and $\cWcl^i(z)$ by
\be
{\cphi} :=  
z^{-1}
\vec\alpha 
+
\sqrt\beta
\sum_{n> 0}\sum_{b=1}^{\nn -1} 
\left(
z^{n-1} 
{
{\vec\Lambda^b}{} }
p^{(b)}_n 
+
z^{-n-1}
{
{\vec\alpha^b}{} }
{n\over \beta}\deldel{ p^{(b)}_n }
\right)
,
\ee
\be
\sum_{i=0}^\nn \cWcl^i(z)(\alpha_0 \partial_z)^{\nn-i}
:=
\NP 
(\alpha_0\partial_z +  {\ch 1{ }} )
(\alpha_0\partial_z +  {\ch 2{ }} )
\cdots
(\alpha_0\partial_z +  {\ch {\nn}{ }} )
\NP
\ee
and $\cWcl^i(z)=:\sum_{n\in\bZ} {\cWcl^i}_n z^{-n}$.
Similarly, ${\tcphi}$ and $\tcWcl^i(z)$ are defined by replacing
$ {n\over \beta}\deldel{ p^{(b)}_n }$
with
$\sum_{k=1}^{r_j} (z_k^{b})^n $.
Then we have the {\W} constraint 
${\cWcl^a}_n Z_\nn = 0$
and the loop equation 
$\CF{ \tcWcl^a{}_n }= 0$ for $n>0$.
%
The quantum spectral curve \eqref{eq:SpectralCurve} is now
\be
\CF{
(\alpha_0\partial_z + {\tch 1{ }} )
(\alpha_0\partial_z + {\tch 2{ }} )
\cdots
(\alpha_0\partial_z + {\tch {\nn}{ }} )
}
=0.
\ee
%
%
%
%
For large $r_a$ this 
reduces to 
\eqref{eq:LargeRSpectralCurve}
with $R=0$.
%
%
%
%
When
$
p^{(a)}_n
= 
\sum_{j=1}^{M_i} (x^{(a)}_j)^n 
+
(\gamma^{(a)}y^{(a)})^n 
,
$
\be
V_\nn
=
\prod_{a=1}^{\nn -1} 
\Exp{
\sqrt\beta
\sum_{n>0}
{1\over n}
{\bosonIP{\vec\Lambda^a}{\vec a_n}{} } 
\left\{
\sum_{j=1}^{M_i} 
(x^{(a)}_j)^n 
+
(\gamma^{(a)}y^{(a)})^n 
\right\}
}
\ee
is the positive mode part of 
$\VO^a_{\gamma^{(a)}}(1/y^{(a)})$ 
and
$\VO^a_+(1/x^{(a)}_j)$.%
\footnote{
The Toda theory/\cW-gravity duality is discussed in \cite{rf:Bonelli-Tanzini}.
}{ }

\subsection{$\nn =2$ case} 


When $\nn =2$, i.e., the Virasoro case,
%
%
%
${\vec\Lambda^1} =  {\vec h^1} = -{\vec h^2}$,
${\vec\alpha^1} = 2 {\vec h^1}$,
${\bosonIP{\vec h^1}{\vec h^1}{}} = A^{1,1} = 1/2$,
$B^{1,1} = 1$ and 
$C^{1,1}= 2$.
Let $p_n:=p_n^{(1)}$ then
the partition function $Z_2 $ is now
\be
Z_2(p) 
=
\oint \prod_{j=1}^{r} {dz_j\over \twoPiI z_j}
z_j^{-s}
\Exp{\beta\sum_{n>0}{1\over n}
z_j^n p_n }
\cdot
\qWDelta(z)
=
{ 
r!\qWPIPbk{ s^r }r
\over 
\PIPbk{ s^r }
}
P_{(s^r)}({\xp p}).
\ee
%
As \eqref{eq:qGMMVir}
the partition function 
$Z_2(p)$ substituting 
$p_n
=\sum_i x_i^n
+(\gamma y)^n$
is
\be
{ 
Z_2 \left(\sum_i x_i+{\gamma}y\right)
\over 
Z_2\left({\gamma}y\right)
}
=
\betaHyperGeo 21{\beta}{-s,r\beta}{1-s+(r-1-\gamma)\beta}{{x\over y}}.
\label{eq:betaGMMVir}%
\ee
Here
$\betaHyperGeo 21{\beta}{a,b}cx$
is the multivariate hypergeometric function 
\be
\betaHyperGeo 21{\beta}{a,b}cx 
:=
\sum_{\lambda \atop \ell(\lambda)\leq M} 
P_\lambda(x;\beta) 
\prod_{(i,j)\in\lambda} {
(a+{j-1}+(1-i)\beta)
(b+{j-1}+(1-i)\beta)
\over
(c+{j-1}+(1-i)\beta)
(\lambda_i-j+1+(\lambda\dualvee_j-i)\beta)
}.
\ee
If we substitute
$ \beta^n p_n
=(-1)^{n-1}
(\sum_i x_i^n
+(\gamma y)^n
)$
to $Z_2(p)$ then the right hand side 
$\betaHyperGeo 21 {\beta} {a,b}c{{x \over y}}$
of \eqref{eq:betaGMMVir}
is replaced by 
$\betaHyperGeo 21 {1/\beta} {-a,-b}{-c}{{x \over y}}$.

\subsection{Four-dimensional Nekrasov partition function} 


Let $a=(a_1,\cdots,a_\nn )$ and 
$m=(m_1,\cdots,m_{2\nn})$
be sets of complex parameters.
$m_k$ corresponds to the mass of the fundamental matter.
Then by
\eqref{eq:NekFactor} and 
\eqref{eq:NekNon},
\ba
Z^{\rm inst}(a)
&:=&
\sum_{\{\lambda_i\}}
\prod_{i=1}^\nn
(-\Lambda^2)^{\nn|\lambda_i|}
{
\prod_{k=1}^{2\nn}
N_{\lambda_i\bullet}( {a_i+m_k\over \sqrt\beta} -{\alpha_0/ 2})
\over
\prod_{j=1}^\nn
N_{\lambda_i\lambda_j}( {a_i-a_j\over \sqrt\beta} )
}
,
\qquad
\alpha_0 := \sqrt\beta-{1\over \sqrt\beta},
\cr
N_{\lambda\mu}(a)
\!:\!&=&
N_{\lambda\mu}(a;\sqrt\beta)
:=
(-1)^{|\lambda|+|\mu|}
\cr
&&\hskip-36pt
\times
\prod_{(i,j)\in\lambda} 
\left( a+{ \lambda_i-j \over\sqrt\beta } +\sqrt\beta({\mu\dualvee_j-i+1}) \right)
\prod_{(i,j)\in\mu } 
\left( a-{\mu_i-j+1\over \sqrt\beta } -\sqrt\beta({\lambda\dualvee_j-i  }) \right)
,~~~~~~
\ea
which 
satisfies
\ba
N_{\lambda\mu}\left(a-{\alpha_0\over 2};\sqrt\beta\right)
&=&
N_{\mu\lambda}\left(a+{\alpha_0\over 2};{-\sqrt\beta}\right)
=
N_{\mu\dualvee\lambda\dualvee}\left(a+{\alpha_0\over 2};{1\over\sqrt\beta}\right)
\cr
&=&
(-1)^{|\lambda|+|\mu|}
N_{\mu\lambda}\left(-a-{\alpha_0\over 2};\sqrt\beta\right)
.
\ea

%
When $\nn =2$,
$Z^{\rm inst}$
coincides with the $M=1$ case of the partition function $Z_2$
\be
Z^{\rm inst}
(-m_1+{\textstyle{\beta-1\over 2}},-m_2+{\textstyle{3\beta-1\over 2}})
=
\betaHyperGeo 21 {}
{
{a_2+ m_3+{1-\beta\over 2} },
{a_2+ m_4+{1-\beta\over 2} }
}
{
{a_2- a_1+1-\beta }
}
{
{\Lambda^4}
}
=
{ Z_2\left(x+{\gamma}y\right) \over Z_2\left({\gamma}y\right) }
\label{eq:betaHyperNek2q}%
\ee
with
${s} = { -a_2-m_3 -{\textstyle {1-\beta\over 2}}}$,
${r}\beta = { a_2+m_4 +{\textstyle {1-\beta\over 2}}}$,
$\gamma\beta = {a_1 + a_2+ {m_3}+ {m_4}+1-\beta }$
and
${x/ y} = 
{\Lambda^4}$.
%
Similarly, $Z^{\rm inst}
(-m_1+{\beta-1\over 2},-m_2+{\beta-3\over 2})$
is given by changing the sign of the variables $a$, $b$ and $c$ of 
$\betaHyperGeo 21 {} {a,b}c{\Lambda^4}$
in \eqref{eq:betaHyperNek2q}.


\subsection{Jack polynomial}


%
Finally the Jack polynomials $P_\lambda(x):=P_\lambda(x;\beta)$ 
are defined by
\ba
&& 
H P_\lambda(x) =\varepsilon_\lambda P_\lambda(x),
\qquad
\varepsilon_\lambda := 
\sum_{i=1}^\mm \lambda_i(\lambda_i+(\mm+1-2i)\beta),
\cr
&&  
H:=
  \sum_{i=1}^\mm D_i^2
  +\beta\sum_{i<j}
  \frac{x_i+x_j}{x_i-x_j}(D_i-D_j),
\qquad
D_x := x{\partial\over\partial x}
\ea
with a normalization condition
$P_\lambda(x) = 
x_1^{\lambda_1} 
x_2^{\lambda_2} 
\cdots
x_\mm^{\lambda_\mm}
+\cdots$.
%
%
Inner products 
$\IP fg{q,t}$ and 
$\MacIP fg\mm$
are the same with 
\eqref{eq:InnerProduct} and \eqref{eq:AnotherInnerProduct}, respectively, 
but
$p_n^* := {n \over\beta} {\partial \over \partial p_n}$
and 
$
\MacDelta(x) 
:=
\prod_{i\neq j}^\mm
\Exp{ -\beta\sum_{n>0}
{x_j^n/n x_i^n}}
=
\prod_{i\neq j}^\mm
 ( 1-x_j/ x_i )^\beta
=
 (-1)^{{r(r-1)\over 2}\beta}
\qWDelta(x)
$.
%
%
The inner products of Jack polynomials are given by 
\ba
\IP{ P_\lambda }{ P_\mu }{q,t}
&=&
\delta_{\lambda,\mu}
\PIP{\lambda},
\qquad
\PIP{\lambda}
:=
\prod_{(i,j)\in\lambda}
{
{\lambda_i-j+1} +({\lambda\dualvee_j-i})\beta
\over 
{\lambda_i-j  } +({\lambda\dualvee_j-i+1})\beta
},
\\
\MacIP{ P_\lambda }{ P_\mu }\mm
&=& 
\delta_{\lambda,\mu}
\MacPIP{\lambda}\mm
,
\quad
{
\MacPIP{\lambda}\mm
\over 
\PIP{\lambda}
}
:=
\prod_{(i,j)\in\lambda}
{{j-1}+({\mm-i+1})\beta \over {j}+({\mm-i})\beta }
\prod_{k=1}^\mm
{\Gamma(k\beta) \over \Gamma(\beta)\Gamma((k-1)\beta+1)}.
~~~~~~
\ea
%
%
%
%
The specialization 
$p_n := \gamma$ with $\gamma\in\bC$
is
\be
P_\lambda\left({\xp\gamma}\right) =
\prod_{(i,j)\in\lambda}
{{j-1}+(\gamma+1-i)\beta \over {\lambda_i-j} +({\lambda\dualvee_j-i+1})\beta }.
\ee
%
We have the following integral representation of the Jack polynomial  
\cite{rf:Jack}
\ba
  P_{\lambda}(x;\beta)
&=&
{\qWC\lambda +}
\langle \alpha_{r,s}^+|\Exp{-\sqrt\beta\sum_{n>0}{{\h 1n}\over n}\sum_{i=1}^Mx_i^n}
|\chi_{r,s}^+\rangle,
\quad
{\qWC\lambda +}
:=
  \prod_{a=1}^{\nn-1}
{
(-1)^{{r^{(a)}(r^{(a)}-1)\over 2}\beta}
\PIP{ \lambda^{(a)} }
\over
r_a!
\MacPIP{ \lambda^{(a)} }{r_a}
}
\cr
P_{\lambda\dualvee}\left(-x;\beta\right)
&=&
{\qWC\lambda -}
\langle \alpha_{r,s}^-|\Exp{-{\sqrt\beta}
\sum_{n>0}{{\h 1n}\over n}\sum_{i=1}^Mx_i^n}
|\chi_{r,s}^-\rangle
,\quad
{\qWC\lambda -}
:=
\omega_-\omega_+
{ {\qWC\lambda +}\over \PIP{ \lambda } }
\ea
with $z^\nn_i:=0$.



\newpage
\Appendix{\appNotation}{Notation}


Here we list up the notation of bosons.

\noindent
${\h in}$ and ${\Qh i}$ ($i=1,2,\cdots,\nn $)
: Fundamental (weight) bosons. 
$\sum_{i=1}^\nn p^{in} {\h in} = \sum_{i=1}^\nn {\Qh i} =0$.

\noindent
${\Lam an}$ and ${\QLa a}$ ($a=1,2,\cdots,\nn -1$)
: Weight bosons. 
${\Lam 0n} = {\Lam Nn} = {\QLa 0} =  {\QLa \nn } = 0$. 

\noindent
${\al an}$ and ${\Qal a}$ ($a=1,2,\cdots,\nn -1$)
: Root bosons.

\subsection{Relations}

$$
\begin{array}{c c c c c c c}

\hline
\rule[-17pt]{0pt}{42pt} 
&& {\h jn}
&& {\Lam bn}
&& {\al bn}
\cr

\hline
&&&&&&\cr

{\h in} 
&&
&=& {\ds
\sum_{b=1}^{\nn -1} {\ThLambda ib{p^n}}{\Lam bn}
}&=& {\ds
\sum_{b=1}^{\nn -1} {\Thalpha ib{p^n}}{\al bn},
}\cr

{\Qh i} 
&&
&=& 
{\QLa i}-{\QLa {i-1}}
&=& {\ds 
\left( \sum_{b=i}^{\nn -1} - \sum_{b=1}^{\nn -1} {b\over \nn } \right) 
{\Qal b},
}\cr

&&&&&&\cr
\hline
&&&&&&\cr

{\Lam an} 
&:=& {\ds 
\sum_{b=1}^{\nn -1} {\TLambdah ab{p^n}}{\h bn}
}&&
&=& {\ds
\sum_{b=1}^{\nn -1} {\TLambdaalpha ab{p^n}}{\al bn},
}\cr

{\QLa a} 
&=& {\ds  
\sum_{b=1}^a {\Qh b} 
}&&
&=& {\ds  
\left( \sum_{b=1}^a b{\nn -a\over \nn } 
+ \sum_{b=a+1}^{\nn -1} a{\nn -b\over \nn } \right)
{\Qal b},
}\cr

&&&&&&\cr
\hline
&&&&&&\cr

{\al an} 
&:=& 
{\h an}-{\h {a+1}n}
&=& {\ds
\sum_{b=1}^{\nn -1} {\TalphaLambda ab{p^n}}{\Lam bn},
}
&& 
\cr

{\Qal a} 
&=& 
{\Qh a}-{\Qh {a+1}}
&=& 
-{\QLa {a-1}} + 2{\QLa a} - {\QLa {a+1}}.
&&
\cr

&&&&&&\cr
\hline

\end{array}
$$

Here
\ba
{\Thalpha ibp}
&:=&
{ { \qpint{ \nn \theta(i\leq b) - b }p } \over { \qpint \nn p } }
p^{{ b-\nn \theta(i>b) \over 2 }},
\hskip26pt
{\Talphah abp}
=
\ddelta_{a,b}-\ddelta_{a+1,b}+p^{b-\nn }\ddelta_{a,\nn -1},
\cr
{\ThLambda ibp} 
&:=&
p^{\ha}\delta_{i,b} - p^{-\ha}\delta_{i-1,b},
\hskip73pt
{\TLambdah abp}
=
p^{b-a-\ha} \theta(a\geq b),
\cr
{\TalphaLambda abp}
&:=&
{\qpint 2p}\ddelta_{a,b} 
- p^{-\ha}\ddelta_{a-1,b}
- p^{\ha}\ddelta_{a+1,b},
\quad
{\TLambdaalpha abp}
=
{ { \qpint{ \min (a,b) }p }{ \qpint{ \nn -\max (a,b) }p } 
\over { \qpint \nn p } }
p^{{b-a\over 2}}.
\cr &&
\ea

\subsection{Commutation relations between bosons}
\vskip12pt

$$
[X_n^i,Y_m^j]=
{1\over n}(q^{{n\over 2}}-q^{-{n\over 2}})(t^{{n\over 2}}-t^{-{n\over 2}})
\ddelta_{n+m,0} Z_n^{i,j}.
$$
$$
\begin{array}{c | c c c}

\hline
\rule[-17pt]{0pt}{42pt} 
~~ X_n^i\backslash Y_{-n}^j ~~
& {\h j{-n}}
& {\Lam b{-n}}
& {\al b{-n}}
\cr

\hline
&&&\cr

{\h in} 
&{\ds
~~{ {\qpint{\delta_{ij}\nn -1}{p^n}} \over {\qpint{\nn }{p^n}} } 
p^{{n\over 2} \nn {\rm sgn}(j-i)}
}& {\ds
{\Thalpha ib{p^n}}
}& {\ds
{\ThLambda ib{p^n}}
}\cr

&&&\cr

{\Lam an} 
& {\ds 
{\Thalpha ja{p^{-n}}}
}&{\ds
{\TLambdaalpha ab{p^n}}
}& {\ds
\delta_{a,b}
}\cr

&&&\cr

{\al an} 
& 
{\ThLambda ja{p^{-n}}}
& {\ds
\delta_{a,b}
}& 
{\TalphaLambda ab{p^n}}
\cr

&&&\cr
\hline

\end{array}
$$

\vskip24pt
\subsection{Commutation relations between boson zero modes}
\vskip12pt

$$
[X_n^i,Q^j]= \ddelta_{n,0} Z^{i,j}.
$$
$$
\begin{array}{c | c c c}

\hline
\rule[-17pt]{0pt}{42pt} 
~~ X_0^i\backslash Q^j ~~
& {\Qh j} 
& {\QLa b} 
& {\Qal b} 
\cr

\hline
&&&\cr

{\h i0}
&{\ds
\delta_{i,j}-{1\over \nn }
}&{\ds 
\theta(i\leq b) - {b\over \nn }
}&{\ds 
\delta_{i,b}-\delta_{i-1,b}
}\cr

&&&\cr

{\Lam a0}
& {\ds ~~ 
\theta(j\leq a) - {a\over \nn }
}&{\ds ~~~~
\min (a,b)\left( 1-{\max (a,b)\over \nn } \right)
}& {\ds  
\delta_{a,b}
}\cr

&&&\cr

{\al a0}
& 
\delta_{a,j}-\delta_{a+1,j}
& 
\delta_{a,b}
&
2\delta_{a,b} -\delta_{a-1,b} - \delta_{a+1,b}
\cr

&&&\cr
\hline

\end{array}
$$




\end{document}